\documentclass[11pt]{article}%
\usepackage{amssymb}
\usepackage{graphicx}
\usepackage{amsmath}
\usepackage{amsfonts}%
\providecommand{\U}[1]{\protect\rule{.1in}{.1in}}
\ifx\pdfoutput\relax\let\pdfoutput=\undefined\fi
\newcount\msipdfoutput
\ifx\pdfoutput\undefined\else
\ifcase\pdfoutput\else
\ifx\paperwidth\undefined\else
\ifdim\paperheight=0pt\relax\else\pdfpageheight\paperheight\fi
\ifdim\paperwidth=0pt\relax\else\pdfpagewidth\paperwidth\fi
\fi\fi\fi
\begin{document}

\title{On Some Theoretical Aspects of Functional Data Analysis of
Parametrized Curves in}
\author{Bernard Colin\\D\'{e}partement de Math\'{e}matiques\\Universit\'{e} de Sherbrooke, Qu\'{e}bec, Canada}

\begin{center}
{\Large On Some Theoretical Aspects of Functional Data Analysis of
Parametrized Curves in }$%
\mathbb{R}
^{p}$

\bigskip

{\large Bernard Colin}

{\large D\'{e}partement de Math\'{e}matiques}

{\large Universit\'{e} de Sherbrooke, Qu\'{e}bec, Canada}

bernard.colin@usherbrooke.ca

\bigskip

\end{center}

\textbf{Abstract} In this paper, one provides a comprehensive mathematical and practical
overview of Functional Data Analysis (FDA) specifically applied to
parametrized curves in $\mathbb{R}^{p}$. One observes that curves depending on continuously parameter are naturally
present across many fields, such as, for example, monitoring child
development in pediatrics, assessing meteorological phenomena, analyzing
financial portfolios, tracking neurological functions, and mapping geographic
pollution levels. One considers a formal theoretical framework given by a
Cartesian product of real separable Hilbert spaces, to model these curves
mathematically. To bridge the gap between discrete experimental measurements
subject to errors, and smooth continuous functions, the paper details the
essential phase of data smoothing and fitting and sets its mathematical
formulation via of the Ordinary or Penalized Least Squares 
criteria which, for the later, using the Sobolev spaces framework, incorporates a smoothing parameter $\lambda$
and differential operators to prevent erratic geometric behaviors by
penalizing excessive curve roughness.
Furthermore, one extends the usual Principal Component Analysis to its
functional counterpart in $\mathbb{R}^{3}$, using the calculus of variations and the Euler-Lagrange multiplier theorem
in order to find eigenfunctions and eigenvalues of the covariance operator to
exhibit the optimal decomposition of the spatial variance and finally, one
highlights the geometric advantages of FDA over traditional multivariate data
analysis, emphasizing its unique capacity to capture critical differential
features as velocity, acceleration, curvature, and torsion.

\section{Introduction}

In the context of studying a given phenomenon, one frequently possesses a set of
observations on which measurements are taken. These measurements
relate to a set of real variables, which are assumed to be continuously
differentiable up to a certain order with respect to a real parameter such as time. This is common, for example, in temporal phenomena or stochastic processes,
but the parameter can also represent a rate, an index, a percentage, a length, a
temperature, or a pressure.

\bigskip

Thus, in a longitudinal study involving a cohort of children of the same sex
followed from age 2 to age 18, anthropometric data relating to variables are
recorded for each subject at predetermined times $:$ $t_{0}=0<t_{1}<$ $t_{2}...<t_{k}\leq T$. These variables may
include height, body mass, torso height, limb length and circumference, and pelvis,
chest, neck, and head circumferences, as well as chest cavity volume. Each
variable is then considered an unknown real function of time, assumed to be
continuously differentiable up to a given order on the time interval, and for which
observed values are available. These observed values are, moreover,
subject to errors due, among other things, to the imprecision of the measuring
instruments.

\bigskip

Given a set of meteorological stations distributed over a certain territory, the
average daily values of variables are recorded. These variables include
temperature, precipitation height, prevailing wind speed and direction, as well as
levels of, sulfur, and various pollutants. Thus, over a duration of six months
or a year, a dataset is available for each meteorological station, allowing the
visualization of the station's trajectory in $\mathbb{R}^{p}$ at selected time intervals.

\bigskip

Consider financial institutions offering their clients a set of financial products.
Among these is a portfolio consisting of mutual funds, identical from one institution
to another, whose composition across the different funds is managed by an account manager
responsible for growing the portfolio. Following fluctuations in stock market
indices and at random times, the account manager modifies, according to their institution's
investment policy, the percentage of the mutual funds making up the portfolio to
optimize its short-term return. One can thus track over time the trajectories
in $\mathbb{R}^{p}$ of a given portfolio, reflecting the behavior of the institutions according to their
risk aversion.

\bigskip

In the field of health sciences, techniques such as MRI, tomography, and CT scans
provide reconstructed images of any part of the human body. In
neuroscience, for instance, images of white matter fiber tracts in a specific part of
the brain are obtained, from which a representation in $\mathbb{R}^{3}$ can easily be deduced. Using this geometric representation, one can then compare white matter fiber
tracts located at the same position in the brain between patients with
neurodegenerative diseases and the others. Similarly, in general or
sports medicine, the same techniques can be used to observe fiber tracts of a
given muscle in a group of patients or athletes to study their shape in the case of
injury or atrophy resulting from wounds or accidents.

\bigskip

At each element of a set of geographic locations $L_{1},L_{2},...,L_{n}$, which may contain densely
populated urban areas, rural zones, or semi-desert and desert regions,
measurements are taken using radiosondes. These radiosondes are
positioned vertically above each location and operate at various predetermined
altitudes $h_{1}%
<h_{2}<...<h_{k}$, which are identical for each location. They measure the
concentrations $x_{1}%
(h_{i}),x_{2}(h_{i}),...,x_{p}(h_{i})$ $i=1,2...,k$, evaluated in \textit{ppm} per unit volume of air, of specific atmospheric
pollutants. Considering the altitude $h$ $(0<h<H)$ as a parameter and assuming that the
measurements were taken at each location on the same date, one obtains a set of
measurements $x_{1j}(h_{i}),x_{2j}(h_{i}),...,x_{pj}(h_{i})$
$i=1,2...,k$ $;$ $j=1,2,...,n$, allowing the representation in $\mathbb{R}^{p}$ of the trajectory of each
considered location, illustrating the local pollution phenomenon as a function of
altitude. For a given level surface, one can also be interested in the relative
position of geographic locations in terms of pollutants present in the atmosphere at
a specific altitude. Furthermore, nothing prevents identical measurements of
the same pollutants from being taken at the same geographic locations as a
function of time $t$ $(0<t<T)$. In this case, the parameter space is a compact subset of $\mathbb{R}^{2}$,
and the trajectory of a given location can be assimilated to a 2-dimensional
sub-manifold of $\mathbb{R}^{p}$.

\bigskip

In what follows, one focuses on functional data analysis when the data consist of
parametrized curves in $\mathbb{R}^{p}$ (in the vast majority of applications, $p=3$). More
formally, one denotes by:

\[
\gamma(t)=(x_{1}(t),x_{2}(t),...,x_{p}(t))
\]
a function of class $\mathcal{C}_{I}^{r}$, defined on an interval
$I$ of $\mathbb{R}$ and taking values in $\mathbb{R}^{p}$. Its
graphical representation is called a curve, which will also be designated, by abuse
of notation, by the symbol $\gamma$. In practice, these functions, which are assumed
to be continuous and differentiable up to order $r$ with respect to the parameter $t$ in $I
$, describe observable phenomena for which the analytical expression
of $\gamma$ generally remains unknown. Despite this reality, the data analyst has at
their disposal, for a finite number $t_{1},t_{2},...,t_{k}$ values of $t$, not the exact values
$\gamma(t_{i})$ of the function $\gamma(t)$ for $i=1,2,...,k$, but observed or measured values $\hat{\gamma}(t_{i})$ which constitute approximations
of the theoretical values $\gamma(t_{i})$. At this stage, it becomes clear that if one wishes to
benefit from the continuity and differentiability assumptions of the considered
phenomenon, the natural option consists of replacing the discrete data by a
single datum represented by a function (or curve) $\gamma(t)$ in $\mathcal{C}%
_{I}^{r}$ that is as close as possible to
the observations, in a sense that remains to be defined. This represents one
of the challenges of data analysis, and this phase is called \textquotedblleft
smoothing\textquotedblright\ or \textquotedblleft
fitting\textquotedblright.

\bigskip

A second major challenge in the present framework is to generalize
multidimensional data analysis, where observations are vectors in $\mathbb{R}^{p}$, to cases
where observations are functions $\gamma(t)$ belonging to infinite-dimensional spaces.
Classical models such as principal component analysis, canonical analysis,
discriminant analysis, supervised or unsupervised classification, regression, linear
models, and ANOVA are evoked here.
In this context, the data analyst is faced with a finite number $\gamma_{1}(t),\gamma_{2}(t),...,\gamma_{n}(t)$ of independent
observations originating from the same observable phenomenon
$\gamma(t)$. Using smoothing techniques, the analyst will associate a unique curve with each
observation according to the principle described above, in order to constitute a new
set of $n$ observations on which the analys
is will be conducted.

\bigskip

This being said, the generalization of classical multidimensional data analysis
models to the case where functional data are families of parametrized curves in $\mathbb{R}^{p}$, constitutes the third challenge. This requires the introduction of a mathematical theoretical framework that allows the desired generalizations via a
rigorous approach. To this end, and drawing inspiration from the
mathematical framework of multidimensional data analysis, it is clear that both the
theory and the resulting calculations must rely on a formal description of the
ambient space of the objects under consideration, curves, in this case, using the notions of vector spaces, inner products, norms, orthogonality, and orthonormal bases, to name just a few

\section{Theoretical Framework}

\subsection{Reviews}

Let $\mathcal{H}_{1},\mathcal{H}_{2},...,\mathcal{H}_{p}$, be $p$ real separable \textit{Hilbert} spaces, equipped with their respective inner products denoted by, $\langle.,.\rangle_{\mathcal{H}_{1}},\langle
.,.\rangle_{\mathcal{H}_{2}},...,\langle.,.\rangle_{\mathcal{H}_{p}}$ and let $\xi_{i}=\{\xi_{ij_{i}}\}_{j_{i}=1}^{\infty}:$ $i=1,2,...,p$, denote the corresponding orthonormal \textit{Hilbertian} bases.
One considers the Cartesian product $\mathcal{H}
$ of the spaces $\mathcal{H}_{1},\mathcal{H}_{2},...,\mathcal{H}_{p}$ defined
by:
\[
\mathcal{H=}\times_{i=1}^{p}\mathcal{H}_{i}=\{x=(x_{1},x_{2},...,x_{i}%
,...x_{p})\}
\]
where $x_{i}$ belong to $\mathcal{H}_{i}$ for all $i=1,2,...,p$, and which is endowed with the inner product $\langle\cdot,\cdot\rangle
_{\mathcal{H}}$ defined by:
\[
\langle x,y\rangle_{\mathcal{H}}=%
{\textstyle\sum\nolimits_{i=1}^{p}}
\langle x_{i},y_{i}\rangle_{\mathcal{H}_{i}}%
\]
for all $x=(x_{1},x_{2},...,x_{p})$ and $y=(y_{1},y_{2},...,y_{p})$
belonging to $\mathcal{H}$. It is well known in this case (see L. Debnath and P.Mikusiński (8)) that $\mathcal{H}$ is a \textit{Hilbert} space, referred to as \textquotedblleft product space\textquotedblright. In
particular, in the case of parametrized curves defined on an interval $\left[  0,T\right]  $ of $\mathbb{R}$ with values in $\mathbb{R}^{p}$, it is standard to consider that for each $i=1,2,...,p$,
each \textit{Hilbert} space $\mathcal{H}_{i}$ is a space of
square-integrable functions, whose inner product is defined by: 
\[
\forall x_{i},y_{i}\in\mathcal{H}_{i}:\langle x_{i},y_{i}\rangle
_{\mathcal{H}_{i}}=%
{\textstyle\int\nolimits_{0}^{T}}
x_{i}(t)y_{i}(t)dt\text{ , \ }%
\]
which implies that:
\[
\forall x_{i}\in\mathcal{H}_{i}:||x_{i}||_{\mathcal{H}_{i}}^{2}=%
{\textstyle\int\nolimits_{0}^{T}}
x_{i}^{2}(t)dt\text{ \ }.\text{ \ }%
\]

\bigskip

For each $i=1,2,...,p$, one considers the sets $\Phi_{i}$ of elements of
 $\mathcal{H}$, defined as follows :
\[
\left\{
\begin{array}
[c]{l}%
\Phi_{1}=\{\varphi_{1j_{1}}=(\xi_{1j_{1}},0,...,0):j_{1}\geq1\}\\
\Phi_{2}=\{\varphi_{2j_{2}}=(0,\xi_{2j_{2}},0,...,0):j_{2}\geq1\}\\
\vdots\\
\Phi_{i}=\{\varphi_{ij_{i}}=(0,..,0,\xi_{ij_{i}},0,..,0):j_{i}\geq1\}\\
\vdots\\
\Phi_{p}=\{\varphi_{pj_{p}}=(0,0,...,0,\xi_{pj_{p}}):j_{p}\geq1\}
\end{array}
\right.
\]
It is easy to verify that each set $\Phi_{i}$ is, for the inner product defined on the product
space $\times_{i=1}^{p}%
\mathcal{H}_{i}$, an orthonormal family of elements of $\mathcal{H}$ and, moreover, the sets $\Phi_{i}$ are, by
construction, pairwise orthogonal. Setting:
\[
\Phi=\cup_{i=1}^{p}\Phi_{i}\text{ \ },\text{ \ }%
\]
one notes that this set constitutes a countable and orthonormal family of elements of $\mathcal{H}$. Denoting the linear subspace of $\mathcal{H}$ spanned by the elements of $\Phi_{i}$, for each $i=1,2,...,p$, by:
\[ 
lin\Phi_{i}%
\]
it follows that:
\[
\mathcal{H=\oplus}_{i=1}^{p}lin\Phi_{i}%
\]
and:
\[
\cap_{i=1}^{p}lin\Phi_{i}=\{0\}
\]

\bigskip

Furthermore, let $x=(x_{1},x_{2},...,x_{i},...x_{p})$ be an arbitrary element of $\mathcal{H}$. For any element of $\varphi_{ij_{i}}$ of
$\Phi$, one has:
\begin{align*}
\langle x,\varphi_{ij_{i}}\rangle_{\mathcal{H}}  &  =\langle(x_{1}%
,x_{2},...,x_{i},...x_{p}),(0,..,0,\xi_{ij_{i}},0,..,0)\rangle_{\mathcal{H}}\\
&  =\langle x_{i},\xi_{ij_{i}}\rangle_{\mathcal{H}_{i}}%
\end{align*}
where $x_{i}$ belongs to $\mathcal{H}_{i}$. Now, if one fixes $i$, one
knows, by virtue of the fact that $\{\xi_{ij_{i}}\}_{j_{i}=1}^{\infty}$ constitutes a complete orthonormal system of $\mathcal{H}_{i}$, that the preceding inner product will be zero for all $j_{i}\geq1$, if and only if $x_{i}=0 $. Varying $i$ from $1$ to $p$, one easily concludes that the only element of $\mathcal{H}$ that is simultaneously orthogonal to all elements of $\Phi$, is none other than the zero element of $\mathcal{H}$. This allows to assert that $\Phi$ is a complete orthonormal system and therefore constitutes a \textit{Hilbertian} base of $\mathcal{H=}\times_{i=1}^{p}\mathcal{H}_{i}$.
In other words, the closure $\overline{\Phi}$ of $\Phi$ is identical to the space $\mathcal{H}$ (hence separable), and any
element of the latter can be uniquely written in the following form:

\[
x=%
{\textstyle\sum\nolimits_{i=1}^{p}}
{\textstyle\sum\nolimits_{j_{i}=1}^{\infty}}
\alpha_{ij_{i}}\varphi_{ij_{i}}%
\]
with:
\[
\alpha_{ij_{i}}=\langle x,\varphi_{ij_{i}}\rangle_{\mathcal{H}}=\langle
x_{i},\xi_{ij_{i}}\rangle_{\mathcal{H}_{i}}%
\]
and where:

\begin{align*}
||x||^{2}  &  =%
{\textstyle\sum\nolimits_{i=1}^{p}}
{\textstyle\sum\nolimits_{j_{i}=1}^{\infty}}
\alpha_{ij_{i}}^{2}\\
&  =%
{\textstyle\sum\nolimits_{i=1}^{p}}
{\textstyle\sum\nolimits_{j_{i}=1}^{\infty}}
\langle x_{i},\xi_{ij_{i}}\rangle_{\mathcal{H}_{i}}^{2}\\
&  =%
{\textstyle\sum\nolimits_{i=1}^{p}}
||x_{i}||_{\mathcal{H}_{i}}^{2}%
\end{align*}

\subsection{Data Smoothing}

Given a real parameter $t$, defined on a bounded interval $I$ of $\mathbb{R}$ which can be
considered, without loss of generality, as the interval $\left[  0,1\right]  $ or, $]0,1[$
(or any other interval more appropriate to the observations), let $\gamma$ be a parametrized curve in $\mathbb{R}%
^{p}$ defined by the equation $\gamma(t)=(x_{1}(t),x_{2}(t),...,x_{p}%
(t))$. One shall assume throughout, that the curve $\gamma$ is continuously differentiable with respect to the variable $t$, up to a given order and, moreover, that it is simple (without self-intersections) and regular ($\gamma^{\prime}(t)\neq0$ for all $t$ in $I)$.

In practice, the parametrized curve $\gamma$ describes the variation, as a function of $t$, of an observed phenomenon
for which however, continuous values are not available but only a finite number of them are recorded at different values the $0<t_{1}<t_{2}<....<t_{k}<T$, of the parameter $t$. It follows that the curve $\gamma$ cannot be known exactly, and the only recourse available to obtain the most precise representation possible, relies on exploiting the
observations to deduce an approximation $\hat{\gamma}$. Naturally, everything just said
about the function $\gamma(t)$ applies to each of the components $x_{1}(t),x_{2}(t),...,x_{p}(t)$. Regarding the approach to adopt for approximating, one possible idea is to perform data interpolation using sufficiently smooth functions to preserve the natural differentiability properties of the observed phenomenon. However, because the measurements taken are subject to errors and the repetition of experiments involves randomness, this would yield a non robust model that depends too heavily
on empirical conditions. Instead, data smoothing consists in determining, within a given family of curves, an optimal curve according to a given criterion.

\bigskip

To this end, we consider the Cartesian product $\mathcal{H=}%
\times_{i=1}^{p}\mathcal{H}_{i}$ of the\textit{\ Hilbert} spaces
$\mathcal{H}_{1} ,\mathcal{H}_{2},...,\mathcal{H}_{p}$. For each $\mathcal{H}_{i}$ one chooses a basis $\xi_{i}=\{\xi_{ij_{i}}\}_{j_{i}%
=1}^{\infty}$ from which we retain the first $q_{i}$
elements $\xi_{i1},\xi_{i2},...,\xi_{iq_{i}}$, and one denotes by $\widehat{\mathcal{H}_{i}}$ the closed subspace of
$\mathcal{H}$ defined by:
\[
\widehat{\mathcal{H}_{i}}=\text{lin}\left(  \{\varphi_{ij_{i}}%
\}:i=1,2...,p\text{ };\text{ }j_{i}=1,2,...,q_{i}\right)
\]
Thus, for any element of $\widehat{\mathcal{H}_{i}}$ one has
:
\begin{align*}
\gamma(t)  &  =(x_{1}(t),x_{2}(t),...,x_{p}(t))\\
&  =%
{\textstyle\sum\nolimits_{j_{i}=1}^{q_{i}}}
\alpha_{ij_{i}}\varphi_{ij_{i}}(t)
\end{align*}
where for all $t$ in $I$, for all $i=1,2,...,p$ and for all
 $j_{i}=1,2,...,q_{i}$, $\varphi_{ij_{i}}(t)=(0,0,..,\xi_{ij_{i}%
}(t),...,0)$. Without loss of generality and to simplify notation, one henceforth assumes that $q_{i}=q$ for all
 $i=1,2,...,p$, modifications required otherwise are immediate.

\subsubsection{Ordinary Least Squares Criterion}

One assumes that the wanted approximation $\hat{\gamma}$ of $\gamma$
belongs to the linear subspace of $\widehat{\mathcal{H}}=\mathcal{\oplus
}_{i=1}^{p}\widehat{\mathcal{H}_{i}}$ of $\mathcal{H}$, meaning that it is of the form:
\[
\hat{\gamma}(t)=%
{\textstyle\sum\nolimits_{i=1}^{p}}
{\textstyle\sum\nolimits_{j_{i}=1}^{q}}
\alpha_{ij_{i}}\varphi_{ij_{i}}(t)
\]
Given that one has $k$ observations denoted by:
\[
\tilde{\gamma}(t_{r})=(\tilde{x}_{1}(t_{r}),\tilde{x}_{2}(t_{r}),...,\tilde
{x}_{p}(t_{r}))\text{ }:r=1,2,...,k\text{\ },\text{ \ }%
\]
the vector $\alpha^{\ast}=(\alpha_{ij_{i}}^{\ast})$ of the components of
$\hat{\gamma}$ in $\widehat{\mathcal{H}}$ will be, in the lest square sense, the solution of the optimization problem :
\[
\underset{\{\alpha_{ij_{i}}\}\in\mathbb{R}^{pq}}{\min}%
{\textstyle\sum\nolimits_{r=1}^{k}}
||\hat{\gamma}(t_{r})-\tilde{\gamma}(t_{r})||^{2}%
\]
that is:
\[
\alpha^{\ast}=(\alpha_{ij_{i}}^{\ast})=\arg\underset{\{\alpha_{ij_{i}}%
\}\in\mathbb{R}^{pq}}{\min}%
{\textstyle\sum\nolimits_{r=1}^{k}}
\left[  ||%
{\textstyle\sum\nolimits_{i=1}^{p}}
{\textstyle\sum\nolimits_{j_{i}=1}^{q}}
\alpha_{ij_{i}}\varphi_{ij_{i}}(t_{r})-\tilde{\gamma}(t_{r})||^{2}\right]
\]
which, due to the pairwise orthogonality of the linear subspaces spanned by $\hat{\Phi}_{1},\hat{\Phi}_{2},...,\hat{\Phi
}_{p}$ where:
\[
\hat{\Phi}_{i}=\text{lin}\left(  \{\varphi_{ij_{i}}\}:\text{ }j_{i}%
=1,2,...,q\right)  \text{, for all }i=1,2,...,p\text{ \ , \ }%
\]
yields:
\[
\alpha^{\ast}=(\alpha_{ij_{i}}^{\ast})=\arg\underset{\{\alpha_{ij_{i}}%
\}\in\mathbb{R}^{pq}}{\min}%
{\textstyle\sum\nolimits_{i=1}^{p}}
\left[
{\textstyle\sum\nolimits_{r=1}^{k}}
||%
{\textstyle\sum\nolimits_{j_{i}=1}^{q}}
\alpha_{ij_{i}}\xi_{ij_{i}}(t_{r})-\tilde{x}_{i}(t_{r})||^{2}\right]
\]

Now:
\[%
{\textstyle\sum\nolimits_{i=1}^{p}}
\left[
{\textstyle\sum\nolimits_{r=1}^{k}}
||%
{\textstyle\sum\nolimits_{j_{i}=1}^{q}}
\alpha_{ij_{i}}\xi_{ij_{i}}(t_{r})-\tilde{x}_{i}(t_{r})||^{2}\right]
\]
will be minimal if and only if each term in the sum is minimal. It follows from
this observation that for each $i=1,2,...,p $,
it suffices to determine, using standard techniques
implemented in the unidimensional case (see, among others: J. Ramsay and B.
Silverman \cite{ram1},\cite{ram2}, F. Ferraty and P. Vieu \cite{fer} , P.
Kokoszka and M. Reimherr \cite{kok} ,T. Hsing and R. Eubank \cite{hsi}), the
vector in $\mathbb{R}^{q}$ whose components $\alpha_{i1}^{\ast}%
,\alpha_{i2}^{\ast},...,\alpha_{iq}^{\ast}$ are, for all $i=1,2,...,p $,
the solution of the optimization problem:
\[
\alpha_{i}^{\ast}=\arg\underset{\alpha_{ij_{i}}\in\mathbb{R}^{q}}{\min}\left[
%
{\textstyle\sum\nolimits_{r=1}^{k}}
||%
{\textstyle\sum\nolimits_{j_{i}=1}^{q}}
\alpha_{ij_{i}}\xi_{ij_{i}}(t_{r})-\tilde{x}_{i}(t_{r})||^{2}\right]
\]
with:
\[
\alpha_{i}^{\ast}=(\alpha_{i1}^{\ast},\alpha_{i2}^{\ast},...,\alpha_{iq}%
^{\ast})^{t}%
\]
Hence:
\[
\alpha^{\ast}=vec(\alpha_{i}^{\ast})\in\mathbb{R}^{pq}%
\]

\subsubsection{Penalized Least Squares Criterion}

One advantage of functional data analysis over classical multidimensional data
analysis lies in its ability to incorporate the differentiability properties of the
functions under consideration. More precisely, if within the theoretical framework above 
one considers a base,
\[
\Phi=\{\varphi_{ij_{i}}:i=1,2,...,p\text{ };j_{i}=1,2,...\}\text{ of
}\mathcal{H=}\times_{i=1}^{p}\mathcal{H}_{i}\text{, }%
\]
then any element of $\gamma$ of $\mathcal{H}$ can be uniquely expressed as:
\[
\gamma=
{\textstyle\sum\nolimits_{i=1}^{p}}
{\textstyle\sum\nolimits_{j_{i}=1}^{\infty}}
\alpha_{ij_{i}}\varphi_{ij_{i}}%
\]
and if $\hat{\gamma}$ is an approximation of $\gamma$ of the form:
\[
\gamma\approx\hat{\gamma}=%
{\textstyle\sum\nolimits_{i=1}^{p}}
{\textstyle\sum\nolimits_{j_{i}=1}^{q}}
\hat{\alpha}_{ij_{i}}\varphi_{ij_{i}}\text{ , }%
\]
it follows that the $k$ order derivative vector $\gamma^{(k)}$ of $\gamma$
is approximated by:
\[
\gamma^{(k)}\approx\hat{\gamma}^{(k)}=%
{\textstyle\sum\nolimits_{i=1}^{p}}
{\textstyle\sum\nolimits_{j_{i}=1}^{q}}
\hat{\alpha}_{ij_{i}}\varphi_{ij_{i}}^{(k)}%
\]
In the case of parametrized curves in $\mathbb{R}^{3}$, for
instance, the geometric characteristics of the curves are associated with the concepts of velocity, acceleration, normal and
 osculating planes, curvature and radius of curvature, torsion, and \textit{Fr\'{e}net} frame. All these quantities involve not only $\gamma(t)$ but also $\gamma^{(1)}(t),\gamma^{(2)}(t)$
and $\gamma^{(3)}(t)$. If one has a sample of size $n$ of such curves associated with $n$ observations of the same
phenomenon, one may want to include these geometric characteristics in the approximation criterion for subsequent statistical analysis.

\bigskip

One then considers the differential operator $\mathcal{D}\gamma$,
assumed to map $\mathcal{H}$ into $\mathcal{H}$, defined by:
\[
\mathcal{D}\gamma=%
{\textstyle\sum\nolimits_{s=0}^{m}}
\alpha_{s}\gamma^{(s)}%
\]
where $\alpha_{0},\alpha_{1},...,\alpha_{m}$ are real functions of the parameter $t$ and where:
\[
\gamma^{(s)}(t)=\frac{d^{s}\gamma(t)}{dt^{s}}:s=0,1,...,m\text{ \ }.
\]
The objective function to be minimized is the written:
\[%
{\textstyle\sum\nolimits_{i=1}^{p}}
\left[
{\textstyle\sum\nolimits_{r=1}^{k}}
||%
{\textstyle\sum\nolimits_{j_{i}=1}^{q}}
\alpha_{ij_{i}}\xi_{ij_{i}}(t_{r})-\tilde{x}_{i}(t_{r})||^{2}\right]
+\lambda||\mathcal{D}\hat{\gamma}||^{2}%
\]
where $\lambda$ is a positive real number called the smoothing parameter.\ If $\lambda=0,$ the
penalty term does not contribute to the sum of squared deviations, and one obtains
the same solution as in ordinary least squares.  However, if $\lambda$ increases, the
penalty term becomes increasingly significant, which penalizes curves that are too wiggly. Finally if
$\lambda$ is large, the solution curve essentially reflects the
behavior of the operator $\mathcal{D}$. In practice, the parameter is determined using
cross-validation methods (simple or generalized) in order to have a balance between
overfitting and underfitting.

Before implementation, the preceding considerations require some theoretical
complements, a short introduction to which is presented below. For a more
complete and detailed discussion on the topic, particularly its use in functional data
analysis, one may consult the works of A. Berlinet and C. Thomas-Agnan \cite{ber}, T. Hsing and R.
Eubank \cite{hsi}, L.Debnath and P. Mikusi\'{n}ski \cite{deb}, and R.A Adams
and P. Fournier \cite{ada}.

\subsubsection{Sobolev Spaces}

From a theoretical point of view, the \textit{Hilbert} spaces $\mathcal{H}%
_{1},\mathcal{H}_{2},...,\mathcal{H}_{p}$ can be arbitrary, but in practice,
spaces $L^{2}$ of square integrable functions are chosen almost exclusively for both geometric reasons (inner product, orthogonality, orthogonal projection, isomorphism with $l^{2}$) and pragmatic reasons (concepts of
variance, covariance, and correlation, essential in probability and statistics).

\bigskip

According to the theoretical framework introduced previously, one assumes that for
each $i=1,2,...,p$, $\mathcal{H}_{i}$ is a space denoted by $L^{2}$, consisting of the equivalence classes with respect
to the \textit{Lebesgue} measure on $\mathbb{R}$, of real-valued, square-integrable functions defined on a bounded interval $I$ de $\mathbb{R}$
$(I=[a,b]$ or $]a,b[:a<b)$. In the context of parametrized curves, the spaces $\mathcal{H}_{i}$ are considered identical to a prototype space $\mathcal{H}$ equipped with the inner product denoted by $\langle\cdot,\cdot\rangle_{\mathcal{H}}$,
so that the product \textit{Hilbert} space is simply $\mathcal{H}^{p}$.

\bigskip

To provide a rigorous framework for the optimization problem stated above, one
considers an open set $\Omega$ of $\mathbb{R} $
and denote by $\mathcal{\tilde{H}}^{m}(\Omega)$ the space of real functions of class $C^{m}(\Omega)$ such that:
\[
\mathcal{D}^{(r)}f\in L^{2}(\Omega)
\]
for all $r$ less or equal to $m$ and where $\mathcal{D}%
^{(r)}$ denotes the differential operator of order $r$. In other words:
\[
\left(  \mathcal{D}^{(r)}f\right)  \mathcal{(}t\mathcal{)=}\frac{d^{r}%
f(t)}{dt^{r}}%
\]
for all $r=0,1,2,...,m$. One then equips $\mathcal{\tilde{H}}^{m}(\Omega)$ with the inner product defined as follows: for any
elements belonging to $\mathcal{\tilde{H}}^{m}(\Omega)$, one sets:
\[
\langle f,g\rangle_{\mathcal{\tilde{H}}^{m}(\Omega)}=%
{\textstyle\int\nolimits_{\Omega}}
{\textstyle\sum\nolimits_{r=0}^{m}}
\left(  \mathcal{D}^{(r)}f\right)  \mathcal{(}t\mathcal{)}\left(
\mathcal{D}^{(r)}g\right)  \mathcal{(}t\mathcal{)}dt
\]
For example, if $\Omega=(a,b)$ one has:
\[
\langle f,g\rangle_{\mathcal{\tilde{H}}^{m}(a,b)}=%
{\textstyle\sum\nolimits_{r=0}^{m}}
\int_{a}^{b}\frac{d^{r}f(t)}{dt^{r}}\frac{d^{r}g(t)}{dt^{r}}dt
\]
The space $\mathcal{\tilde{H}}^{m}(\Omega)$ equipped with the preceding inner product, is not generally a \textit{Hilbert} since it is not complete. Its completion, denoted by $\mathcal{H}^{m}(\Omega)$, is a \textit{Hilbert}
space belonging to a more general class of spaces denoted by $\mathcal{W}^{m,p}(\Omega)$, better known as
\textit{Sobolev} spaces, which are defined within the framework of $L^{p}(\Omega)$ spaces for $1<p<\infty$. In
the present case, one has $\mathcal{H}^{m}(\Omega)=\mathcal{W}^{m,2}(\Omega)$ and thus the differential operator $\mathcal{D}\gamma=%
{\textstyle\sum\nolimits_{s=0}^{m}}
\alpha_{s}\gamma^{(s)}$, introduced in the previous section, is indeed an operator from the space $\mathcal{H}$ into itself. Furthermore, since in the case of spaces $L^{2}(\Omega)$, the space $\mathcal{W}^{m,2}(\Omega)$
is a \textit{Hilbert} space, it follows that the product space:
\[
\mathcal{W(}\Omega\mathcal{)=\times}_{i=1}^{p}\mathcal{W}^{m_{i},2}(\Omega)
\]
will also be a \textit{Hilbert} space thereby allowing the problem to be treated \textquotedblleft component by component\textquotedblright.

\section{From Observations to Functional Data}

In practice, one assumes that the \textit{Hilbert} spaces $\mathcal{H}_{i},$ $i=1,2,...,p$ are $L^{2}$ spaces, defined on the same domain, and their bases may or may not be identical. For a given phenomenon described by the function or curve
$\gamma(t)=(x_{1}(t),x_{2}(t),...,x_{p}(t))$ of $\mathbb{R}^{p}$, one has, for each element of a sample of size $n$ (a weather site, a city, a country, an individual, a patient...), a set of $k$ measurements $\{\tilde{\gamma}(t_{r}%
)=(\tilde{x}_{1}(t_{r}),\tilde{x}_{2}(t_{r}),...,\tilde{x}_{p}(t_{r}))$ $;$
$r=1,2,...,k\}$ of the considered phenomenon, taken at values $0<t_{1}<t_{2}<....<t_{k}<T$ of $t$.

\subsection{Smoothing}

One assumes that for all $i=1,2,...,p$, a \textit{Hilbertian} base $\xi_{i}=\{\xi_{ij_{i}}\}_{j_{i}=1}^{\infty}$ (orthogonal polynomials, splines, wavelets, trigonometric polynomials...) has been choosen for each \textit{Hilbert} $\mathcal{H}_{i}$ space from which one retains the first $q$ elements and one reduces the search of the parametrized curve $\gamma$ that best fit the data, to the closed subset $\widehat{\mathcal{H}}$ of $\mathcal{H}$ defined by:
\[
\widehat{\mathcal{H}}=\text{lin}\left(  \{\varphi_{ij_{i}}\}:i=1,2...,p\text{
};\text{ }j_{i}=1,2,...,q\right)
\]
where:
\[
\varphi_{ij_{i}}=(0,..,0,\xi_{ij_{i}},0,..,0):j_{i}=1,2...,q
\]
and any element $\hat{\gamma}$ of $\widehat{\mathcal{H}}$ can be written in the form:
\[
\hat{\gamma}=%
{\textstyle\sum\nolimits_{i=1}^{p}}
{\textstyle\sum\nolimits_{j_{i}=1}^{q}}
\alpha_{ij_{i}}\varphi_{ij_{i}}%
\]
where $\alpha_{ij_{i}}=\langle\hat{\gamma},\varphi_{ij_{i}}\rangle
_{\mathcal{H}_{i}}$. Referring back to the results in 2.2.1, the squared error between the
observed values $\tilde{\gamma}(t_{r})$ and the predicted values $\hat{\gamma}(t_{r})$ is given by:
\[%
{\textstyle\sum\nolimits_{r=1}^{k}}
||\hat{\gamma}(t_{r})-\tilde{\gamma}(t_{r})||^{2}=%
{\textstyle\sum\nolimits_{i=1}^{p}}
\left[
{\textstyle\sum\nolimits_{r=1}^{k}}
||%
{\textstyle\sum\nolimits_{j_{i}=1}^{q}}
\alpha_{ij_{i}}\xi_{ij_{i}}(t_{r})-\tilde{x}_{i}(t_{r})||^{2}\right]
\]
and the latter will be minimized if and only if, for each $i=1,2,...,p$,
one finds the values $\alpha_{ij_{i}}^{\ast}$  of the coefficients
$\alpha_{ij_{i}}$ $\phi_{i}$, such that the real function
$\alpha_{i1},\alpha_{i2},...,\alpha_{iq}$ of the $q$ variables given by:
\[
\phi_{i}(\alpha_{i1},\alpha_{i2},...,\alpha_{iq})=%
{\textstyle\sum\nolimits_{r=1}^{k}}
||%
{\textstyle\sum\nolimits_{j_{i}=1}^{q}}
\alpha_{ij_{i}}\xi_{ij_{i}}(t_{r})-\tilde{x}_{i}(t_{r})||^{2}\text{ \ },\text{
\ }%
\]
is minimized. To this end, one sets:
\[
\Lambda_{i}=\left[
\begin{array}
[c]{cccc}%
\xi_{i1}(t_{1}) & \xi_{i2}(t_{1}) & \cdots & \xi_{iq}(t_{1})\\
\xi_{i1}(t_{2}) & \xi_{i2}(t_{2}) & \cdots & \xi_{iq}(t_{2})\\
\vdots & \vdots & \vdots & \vdots\\
\xi_{i1}(t_{k}) & \xi_{i2}(t_{k}) & \cdots & \xi_{iq}(t_{k})
\end{array}
\right]  \text{ ; }\alpha_{i}=\{\alpha_{ij_{i}}\}=\left[
\begin{array}
[c]{c}%
\alpha_{i1}\\
\alpha_{i2}\\
\vdots\\
\alpha_{iq}%
\end{array}
\right]  \text{ et }\tilde{x}_{i}=\left[
\begin{array}
[c]{c}%
\tilde{x}_{i}(t_{1})\\
\tilde{x}_{i}(t_{2})\\
\vdots\\
\tilde{x}_{i}(t_{k})
\end{array}
\right]
\]
It then follows that: 
\begin{align*}
\phi_{i}(\alpha_{i1},\alpha_{i2},...,\alpha_{iq})  &  =||\tilde{x}_{i}%
-\Lambda_{i}\alpha_{i}||^{2}=\left(  \tilde{x}_{i}-\Lambda_{i}\alpha
_{i}\right)  ^{t}(\tilde{x}_{i}-\Lambda_{i}\alpha_{i})\\
&  =\alpha_{i}^{t}\Lambda_{i}^{t}\Lambda_{i}\alpha_{i}-2\alpha_{i}^{t}%
\Lambda_{i}^{t}\tilde{x}_{i}+\tilde{x}_{i}^{t}\tilde{x}_{i}%
\end{align*}
Differentiating this expression with respect to the vector $\alpha_{i}$ and setting this derivative
to $0$, one has:
\[
\Lambda_{i}^{t}\Lambda_{i}\alpha_{i}=\Lambda_{i}^{t}\tilde{x}_{i}%
\]
which, assuming that the matrix $\Lambda_{i}^{t}\Lambda_{i}$ is
invertible, which is very often the case, leads to the solution:
\[
\alpha_{i}^{\ast}=(\alpha_{i1}^{\ast},\alpha_{i2}^{\ast},...,\alpha_{iq}%
^{\ast})^{t}=\left[  \Lambda_{i}^{t}\Lambda_{i}\right]  ^{-1}\Lambda_{i}%
^{t}\tilde{x}_{i}%
\]
Consequently, for each $i=1,2,...,p$, the least squares estimator $\hat{x}_{i}(t)$ of the $i^{th}$ coordinate $x_{i}(t)$ of the function $\gamma(t)$, is given by:
\[
\hat{x}_{i}(t)=%
{\textstyle\sum\nolimits_{j_{i}=1}^{q}}
\alpha_{ij_{i}}^{\ast}\xi_{ij_{i}}(t)
\]
or equivalently, denoting by $\xi_{i}(t)$ the vector with components
$\xi_{i1}(t),\xi_{i2}(t),...,\xi_{iq}(t)$, by:
\[
\hat{x}_{i}(t)=\alpha_{i}^{\ast t}\xi_{i}(t)
\]
This result is nothing else that the least-squares estimator of the parameter $\alpha_{i}$ in the regression model
\[
\tilde{x}_{i}=\Lambda_{i}\alpha_{i}+\epsilon
\]
and for which the predicted value is given by:
\[
\hat{x}_{i}=\Lambda_{i}\alpha_{i}^{\ast}=\Lambda_{i}\left[  \Lambda_{i}%
^{t}\Lambda_{i}\right]  ^{-1}\Lambda_{i}^{t}\tilde{x}_{i}%
\]

Denoting by $\Lambda$ the diagonal matrix formed by the diagonal blocks
$\Lambda_{1},\Lambda_{2},...\Lambda_{p}$, by $\tilde{x}$ the vector
$vec(\tilde{x}_{i})$ and by $\alpha^{\ast}$ the vector $vec(\alpha_{i}^{\ast})
$, it is easy to verify that the set of solutions to the optimization
problem is given by the equality:
\[
\alpha^{\ast}=\left[  \Lambda^{t}\Lambda\right]  ^{-1}\Lambda^{t}\tilde{x}%
\]
It will be noted that for each $i=1,2,...,p$, the matrices $\Lambda_{i}\left[
\Lambda_{i}^{t}\Lambda_{i}\right]  ^{-1}\Lambda_{i}^{t}$, as well as the matrix
$\Lambda\left[  \Lambda^{t}\Lambda\right]  ^{-1}\Lambda^{t}$, are idempotent
and are therefore associated with an orthogonal projection operator.
Geometrically speaking, the solution to this optimization problem consists of
orthogonally projecting the vector of observations:%
\[
\tilde{\gamma}=(\tilde{\gamma}(t_{1}),\tilde{\gamma}(t_{2}),...,\tilde{\gamma
}(t_{k}))
\]
onto the subspaces $\widehat{\mathcal{H}}_{i}$ of $\widehat{\mathcal{H}}$,
which are pairwise orthogonal and where  $\widehat{\mathcal{H}}_{i}=$lin$\left(
\{\varphi_{ij_{i}}\}:j_{i}=1,2,...,q\right)  $.

\bigskip

Finally, if, $\gamma_{l}$ denotes the parametrized curve associated with the $l^{th}$ observation, one denotes by $\hat{\gamma}_{l}$ by its least-squares estimator, whose value at any $t$ in $I$ is given by:
\begin{align*}
\hat{\gamma}_{l}(t)  &  =%
{\textstyle\sum\nolimits_{i=1}^{p}}
{\textstyle\sum\nolimits_{j_{i}=1}^{q}}
\alpha_{l,ij_{i}}^{\ast}\varphi_{ij_{i}}(t)\\
&  =(\hat{x}_{l,1}(t),\hat{x}_{l,2}(t),...,\hat{x}_{l,i}(t),...,\hat{x}%
_{l,p}(t))
\end{align*}
where:
\[
\hat{x}_{l,i}(t)=%
{\textstyle\sum\nolimits_{j_{i}=1}^{q}}
\alpha_{l,ij}^{\ast}\xi_{ij_{i}}(t)
\]

\bigskip

Regarding the numerical aspects of finding solutions, the \textit{fda} et \textit{refund} libraries dedicated to functional data analysis, can be found in $R$ \cite{r}. One can consult with profits the manuals by J. Ramsay \textit{et all.} \cite{ram3} and
by P. Kokoszka and M. Reimherr \cite{kok} in which numerous examples are presented
alongside corresponding codes written in the $R$ language.

\subsection{Usual Functional Descriptive Statistics}

One assumes that for each $l=1,2,...,n$, the observed data:
\[
(\tilde{x}_{l,1}(t_{r}),\tilde{x}_{l,2}(t_{r}),...,\tilde{x}_{l,p}%
(t_{r}))\text{ };\text{ }r=1,2,...,k
\]
have been associated with the functional object $\hat{\gamma}_{l}(t)
$ by means of an expansion, for each component, in appropriate bases, possibly with smoothing parameters
$\lambda_{1},\lambda_{2},...,\lambda_{p}$. As in the unidimensional case, one defines the mean curve, the empirical
variance-covariance matrix, and the empirical correlation matrix, denoted
respectively by $\bar{\gamma}(t),\hat{\Sigma}(t),\hat{\Gamma}(t,s)$, as follows:
\[
\left\{
\begin{tabular}
[c]{l}%
$\bar{\gamma}(t)=\frac{1}{n}%
{\textstyle\sum\nolimits_{l=1}^{n}}
\hat{\gamma}_{l}(t)$\\
\\
$\hat{\Sigma}(t)=\frac{1}{n}%
{\textstyle\sum\nolimits_{l=1}^{n}}
(\hat{\gamma}_{l}(t)-\bar{\gamma}(t))(\hat{\gamma}_{l}(t)-\bar{\gamma}%
(t))^{t}$\\
\\
$\hat{\Gamma}(t,s)=\frac{1}{n}%
{\textstyle\sum\nolimits_{l=1}^{n}}
(\hat{\gamma}_{l}(t)-\bar{\gamma}(t))(\hat{\gamma}_{l}(s)-\bar{\gamma}%
(s))^{t}$%
\end{tabular}
\right.
\]

\subsection{Particular Case of $\mathbb{R}^{3}$}

In practice, the ambient space is very often $\mathbb{R}^{3}$, equipped with its standard canonical
basis. 
In this framework, the derivatives of a parametrized curve $\gamma(t)=(x(t),y(t),z(t)))$, are, for every $t$ in $T$, associated with geometric characteristics of the latter that one might wish to take into account during subsequent statistical analyses.

More precisely, if $\gamma(t)$ denotes a parametrized curve in $\mathbb{R}^{3}$ that is simple, regular, and of
class $C_{T}^{r}$ $(r\geq1)$, its successive derivatives
$\gamma^{\prime}(t),\gamma^{\prime\prime}(t),...,\gamma^{(r)}(t),...$ are also parametrized curves which, as the initial curve, can give rise to the same types of analysis.
Among these derivatives, those of order $1,2$ and $3$ play a very important role
regarding the geometry of the curve, and a few elementary reminders concerning
them are provided below. For a comprehensive and detailed account of the
subject, one can consult the works of M.P. do Carmo \cite{car} and K. Tapp
\cite{tap}. Finally, although it is customary, due to the simplification of
expressions and calculations that this choice entails, to use a parametrization as
a function of the arc length, given by:
\[
s(t)=%
{\textstyle\int\nolimits_{0}^{t}}
\left[  x^{\prime}(u)^{2}+y^{\prime}(u)^{2}+z^{\prime}(u)^{2}\right]
^{1/2}du\text{ \ , \ }%
\]
one prefers, due to the natural context of the problem, to use the standard
parametrization.

\subsubsection{Velocity Vector and Normal Plane}

If $\gamma=\gamma(t)$ is a given parametrized curve, the velocity vector, denoted $\gamma^{^{\prime}}(t)$, is by definition the vector with components :
\[
\gamma^{^{\prime}}(t)=(x^{\prime}(t),y^{\prime}(t),z^{\prime}(t))=\frac
{d\gamma(t)}{dt}%
\]

This vector, also denoted by $v(t)$, is tangent to the curve $\gamma$ at the point $\gamma(t)$ and its Euclidean
norm :
\[
||v(t)||=||\gamma^{^{\prime}}(t)||=\left[  x^{\prime}(t)^{2}+y^{\prime}%
(t)^{2}+z^{\prime}(t)^{2}\right]  ^{1/2}%
\]
is called the instantaneous speed. One then denotes by $\sigma(t)$ the unit vector along
the support of the velocity vector $v(t)$ defined by:
\[
\sigma(t)=\frac{v(t)}{||v(t)||}%
\]
Thus, using standard notation, one has:
\[
\mathbf{v}=||v||\mathbf{\sigma}%
\]
where $\mathbf{\sigma}$ denotes the unit vector oriented along the tangent in the direction of
motion, where $\mathbf{v}$ denotes the velocity vector supported by the tangent, and where $v$ denotes the
instantaneous speed. It will be noted, moreover, that the endpoints of the
elements of the vector field $\mathbf{\sigma}$ belong to the unit sphere of $\mathbb{R}^{3}$. Finally, the arc
length of the curve between the values $t_{1}$ and $t_{2}$ is given by:
\[
l(t_{1},t_{2})=%
{\textstyle\int\nolimits_{t_{1}}^{t_{2}}}
||\gamma^{^{\prime}}(t)||dt
\]
It will be noted that if the curve is traversed at a constant speed, i.e., if $||\gamma^{^{\prime}}(t)||=v$, then:
\[
l(t_{1},t_{2})=v(t_{2}-t_{1})
\]
Moreover, if the position vector $\gamma(t)$ has a constant norm $c$
$(||\gamma(t)||=c$ for all $t\in T)$, $\gamma$ is then a curve belonging to the surface of the sphere in $\mathbb{R}^{3}$, centered at the origin and of radius $c$, and the vectors $\gamma(t)$ and $\gamma
^{\prime}(t)$ are orthogonal for all $t$. Finally, the plan $\Pi_{t}$ passing through
the point $\gamma(t)$ and orthogonal to \`{a} $v(t)$, is called the normal plan to the curve
 $\gamma$ at $t$.

\subsubsection{Acceleration vector $a(t)$ and normal $n(t)$}

The acceleration vector, denoted $a(t),$ is by definition the derivative vector of the
velocity vector $v(t)$, that is, the vector defined by:
\[
a(t)=\frac{dv(t)}{dt}=\frac{d^{2}\gamma(t)}{dt^{2}}=\gamma^{\prime\prime}(t)
\]
The vectors $v(t)$ and $a(t)$ locally span at $\gamma(t)$, a plane involved in the definition of the
osculating plane to a curve, and they define two vector fields known as: the velocity
field and the acceleration field.

\bigskip

The tangential component of acceleration is expressed as:
\[
\left\langle a(t),\sigma(t)\right\rangle \sigma(t)
\]
and therefore its orthogonal projection $a_{\Pi_{t}}(t)$ onto the plane $\Pi_{t}$, is given by:
\[
a_{\Pi_{t}}(t)=a(t)-\left\langle a(t),\sigma(t)\right\rangle \sigma(t)
\]
The support of this vector is called the principal normal to the curve
$\gamma$ at $\gamma(t)$ and the vector defined by:
\[
n(t)=\frac{a_{\Pi_{t}}(t)}{||a_{\Pi_{t}}(t)||}%
\]
is called the unit normal vector and is conventionally denoted by $\mathbf{n}$. Finally, if
the curve is traversed at constant speed
($||v(t)||=v$ for all $t\in T$), the velocity and acceleration vectors are orthogonal since:%
\[
\frac{d}{dt}||v(t)||^{2}=0=\frac{d}{dt}\left\langle v(t),v(t)\right\rangle
=2\left\langle v(t),a(t)\right\rangle
\]

\subsubsection{Curvature and Radius of Curvature}

By definition, the curvature $\kappa(t)$ of the curve $\gamma$ at
$\gamma(t)$ is defined by:
\[
\kappa(t)=\frac{||a_{\Pi_{t}}(t)||}{||v(t)||^{2}}%
\]

and:
\[
R(t)=\frac{1}{\kappa(t)}=\frac{||v(t)||^{2}}{||a_{\Pi_{t}}(t)||}%
\]
is called the radius of curvature. Geometrically, this means that a circle of radius $R(t)$, whose center belongs to $\mathbf{n}$, and which is contained in the plane spanned by $\mathbf{\sigma}$ and $\mathbf{n}$, will be tangent to the curve $\gamma$ at $\gamma(t)$.

\subsubsection{Osculating Plane, Binormal, and Torsion}

The plane spanned by the orthogonal unit vectors $\mathbf{\sigma} $
and $\mathbf{n}$ is called the osculating plane, and the normal to the curve perpendicular to the osculating plane, is called
the binormal, with unit vector $\mathbf{b=\sigma\wedge n}$ oriented such that the frame$(\mathbf{\sigma,n,b}),$ known as
\textit{Frénet} frame, has the same orientation as the usual frame $(i,j,k)$.

\bigskip

By definition, the torsion of the curve $\gamma$ at $\gamma(t)$
is given by:
\[
\tau(t)=-\frac{\langle b^{\prime}(t),n(t)\rangle}{||v(t)||}%
\]
and the radius of torsion $T(t)$ is expressed as:
\[
T(t)=\left\vert \frac{1}{\tau(t)}\right\vert
\]

The frame is a moving reference system through which the instantaneous
curvature and torsion at the point $\gamma(t)$ are expressed, and it represents, in some
sense, the geometric signature of the curve
$\gamma$.

All the expressions introduced above can be seen as functional objects and can
therefore give rise, in turn, to various types of functional data analysis.

It is important, furthermore, to specify that the preceding definitions, including those 
of curvature and torsion, are independent of the chosen parametrization, and it is
known moreover (see for example $(4)$) that two curves $\gamma_{1}$ and $\gamma_{2}$ which possess the
same curvatures and the same torsions at every point $\gamma_{1}(t)$ and $\gamma_{2}(t)$ are identical, up to an isometry of
$\mathbb{R}^{3}$. All the preceding expressions involve the derivatives of order 1, 2, and 3 (the third-order derivative appears in the expression $b^{\prime
}(t)$) of the vector $\gamma(t)$) It follows that if, during the smoothing phase, one wishes to exert some
control over the curvature and torsion in order to prevent solutions from behaving somewhat erratically, it would be useful to utilize a penalized least-squares criterion involving the magnitude of the derivatives concerned.

\bigskip

\subsection{Stochastic Framework}

Let $\left(  \Omega,\mathcal{F},\mathbb{P}\right)  $ be a probability space called \textquotedblleft space of observations\textquotedblright,\ such as a set of individuals, patients, cities, weather sites, industries... and let
$T$ be a parameter space. Because it is frequently the case in the applications, this index is called: \textquotedblleft
time\textquotedblright\ and it is assumed to be a positive real number $t$ generally belonging to the interval $\left[  0,T\right] $  or to $\mathbb{R}^{+}$.

Within the theoretical framework introduced previously, one denotes by 
$X(\omega,t);\omega\in\Omega,t\in T$ a mapping from $\left(
\Omega,\mathcal{F},\mathbb{P}\right)  \times T$ to $\mathcal{H}\times\left(
\mathbb{R}^{p},\mathcal{B}(\mathbb{R}^{p})\right) $ and one says that it is a continuous-time multidimensional stochastic
process if, for every $t$ the partial mapping denoted by $X_{t}(\omega)$, is a random vector defined on $\left(  \Omega,\mathcal{F},\mathbb{P}\right)  $ and taking values in $\left(  \mathbb{R}^{p},\mathcal{B}(\mathbb{R}%
^{p})\right)  $ and if for every $\omega$ the partial mapping denoted by:
\[
X(\omega,\cdot)=\gamma(\omega,\cdot)=(x_{1}(\omega,\cdot),x_{2}(\omega
,\cdot),...,x_{p}(\omega,\cdot))
\]
is an element of $\mathcal{H}$ referred to as the \textquotedblleft
trajectory\textquotedblright\ of $\omega$.

This stochastic structure is superimposed onto the geometric structure of the
\textit{Hilbert} space $\mathcal{H}$ and proves unavoidable for studying the asymptotic properties of
estimators, hypothesis testing, and the convergence of various procedures.
Naturally, such a study requires that one chooses realistic hypotheses both for the
probability measure $\mathbb{P}$, and for the laws $\mu_{t}$ of the family $\{X_{t}(\omega)\}$ of random vectors. In
this sense, while accepting the hypothesis that the process is of second-order
seems natural, the same cannot be said regarding its stationarity. We will
not discuss this topic further, prioritizing for the moment the descriptive and
empirical aspects of the subject. Asymptotic aspects will eventually be
studied later, using the framework proposed by D. Bosq (\cite{bos}).

\bigskip

\section{Principal Component Analysis}
\subsection{Statement of the Problem}

Considering the theoretical framework introduced previously one has
a sample of $n$ independent observations of a given phenomenon, described by a
function $\gamma(t)$ of a real parameter $t$, evaluated for each observation at measurement
points $t_{1},t_{2},..,t_{k}$ of $t$ in $T$ which may or may not be identical in both number and
values. In the present case, the measurement points will be considered
identical for all observations. We further assume that this phenomenon is
continuously differentiable with respect to $t$ up to a given order and that with each
observation is associated the functional object $\hat{\gamma}_{l}(t)$, resulting from the fitting or
smoothing of the true unknown function $\gamma_{l}(t)$ as determined in Section $3.1$. The
sample under consideration then consists of the $n$ parameterized curves $\hat{\gamma}_{1}%
(t),\hat{\gamma}_{2}(t),...,\hat{\gamma}_{n}(t)$ with which one associates the usual descriptive statistics: the empirical mean, variance, standard deviation, covariance, and correlation. It can be noted in this
regard that, if one adopts a geometric and empirical approach, it is not necessary to
resort to integral calculus; however, if one wishes to analyze the asymptotic
behavior of the various procedures involved, one must use the \textit{Bochner} integral, which is more straightforward to use when the \textit{Hilbert} spaces considered, are $L^{2}$ spaces.

Although it is not strictly necessary, one assumes for the sake of simplifying the
presentation that all \textit{Hilbert} spaces $\mathcal{H}_{i}$ are for all $i=1,2,...,p$, identical to the 
\textit{Hilbert} space $L_{T}^{2}$, and that one  has selected from each \textit{Hilbertian } base, identical or not, the first $q$ elements $\{\xi_{1j_{1}}\},\{\xi_{2j_{2}}%
\},...,\{\xi_{pj_{p}}\}$ where $j_{i}=1,2,...,q$ for all $i=1,2,...,p$. It follows that the subspace $\widehat{\mathcal{H}}$\ of $\mathcal{H}^{n}$
defined by: %
\[
\widehat{\mathcal{H}}=\text{lin}\left(  \{\varphi_{ij_{i}}\}:i=1,2...,p\text{
};\text{ }j_{i}=1,2,...,q\right)
\]
where:
\[
\varphi_{ij_{i}}=(0,..,0,\xi_{ij_{i}},0,..,0)
\]
is a closed subspace of $\mathcal{H}$ and any element $\gamma$
of the latter can be expressed in the form:
\[
\gamma=%
{\textstyle\sum\nolimits_{i=1}^{p}}
{\textstyle\sum\nolimits_{j_{i}=1}^{q}}
\beta_{ij_{i}}\varphi_{ij_{i}}%
\]
or equivalently:
\[
\gamma=\left(
{\textstyle\sum\nolimits_{j_{1}=1}^{q}}
\beta_{1j_{1}}\xi_{1j_{1}},%
{\textstyle\sum\nolimits_{j_{2}=1}^{q}}
\beta_{2j_{2}}\xi_{2j_{2}},...,%
{\textstyle\sum\nolimits_{j_{i}=1}^{q}}
\beta_{ij_{i}}\xi_{ij_{i}},...,%
{\textstyle\sum\nolimits_{j_{p}=1}^{q}}
\beta_{pj_{p}}\xi_{pj_{p}}\right)
\]
Similarly, for every $l=1,2,...,n$, $\hat{\gamma}_{l}$ can be expressed in the form:
\[
\hat{\gamma}_{l}=\left(
{\textstyle\sum\nolimits_{j_{1}=1}^{q}}
\alpha_{l,1j_{1}}^{\ast}\xi_{1j_{1}},%
{\textstyle\sum\nolimits_{j_{2}=1}^{q}}
\alpha_{l,2j_{2}}^{\ast}\xi_{2j_{2}},...,%
{\textstyle\sum\nolimits_{j_{i}=1}^{q}}
\alpha_{l,ij_{i}}^{\ast}\xi_{ij_{i}},...,%
{\textstyle\sum\nolimits_{j_{p}=1}^{q}}
\alpha_{l,pj_{p}}^{\ast}\xi_{pj_{p}}\right)
\]
In the following, one  assumes that the observations $\hat{\gamma}_{l}$ are assigned strictly positive
weights $p_{l}$ shuch that $\Sigma_{l=1}%
^{n}p_{l}=1$. In practice, and unless stated otherwise, we assume
that $p_{l}=\frac{1}{n}$ for all $l$.

\subsection{Principal Components}

According to the stochastic framework previously introduced, one considers a stochastic
process $X(\omega,t)$ defined on $(\Omega,\mathcal{F},\mathbb{P})\times T$ \`{a} with values in $\mathcal{H}%
\times\mathbb{R}^{p}$, assumed to be of second order $(\mathbb{E(}%
||X(\omega,t)||^{2})<\infty$ for all $t)$ whose mean function is denoted by $\mu(t)=\mathbb{E}(X(t))$ and whose autocovariance function is given by:
\[
\Gamma(t,s)=\mathbb{E}[X(t)X(s)]
\]
Furthermore, one denotes by $\gamma_{1}%
(t),\gamma_{2}(t),...,\gamma_{l}(t),...,\gamma_{n}(t)$, $n$ independent realizations of this process, observed at
values $0<t_{1}%
<t_{2}<...<t_{k}<T$ of the real parameter $t$ which is assumed, without loss of generality, to
belong to a closed bounded interval of the form $\left[  0,T\right]  $. Denoting by
$\hat{\gamma}_{1}(t),\hat{\gamma}_{2}(t),...,\hat{\gamma}_{l}(t),...,\hat
{\gamma}_{n}(t)$ the least-squares approximations of the trajectories $\gamma_{l}(t)$,
$l=1,2,...,n$, one wants to find the element of $C_{1}(t)$ of
$\mathcal{H=}L_{\left[  0,T\right]  }^{2}$, with unit norm ($||C_{1}%
(t)||_{\mathcal{H}}^{2}=1$), which best summarizes the dispersion of the set of the $n$ parametrized curves $\hat{\gamma}_{1}(t),\hat{\gamma}_{2}(t),...,\hat{\gamma
}_{l}(t),...,\hat{\gamma}_{n}(t)$. The solution to this optimization problem, if it exists, will be called the \textquotedblleft
the first principal curve\textquotedblright\ or, more usually, \textquotedblleft first principal component\textquotedblright\ and both terms will be used interchangeably depending on the context. Subsequently, one wants to find a curve
$C_{2}(t)$, with unit norm and orthogonal to $C_{1}(t)$ ($\left\langle
C_{1}(t),C_{2}(t)\right\rangle _{\mathcal{H}}=0$), that best summarizes
the residual dispersion of the $n$ curves, not explained by
$C_{1}(t)$. This solution curve will be called the \textquotedblleft
second principal component\textquotedblright\ and so on until a given
stopping criterion is met.

\bigskip

As an introduction, one presents below the notion of principal curves of a given and finite set of curves, defined on a closed bounded interval of $\mathbb{R}$ and with real values.

\subsubsection{Generalities}

One considers $n$ known real functions $x_{1}(t),x_{2}(t),...,x_{n}(t)$, defined and continuous on a closed and bounded interval $I$ of $\mathbb{R}$, which is assumed to be of the form $\left[0,T\right]$, and belonging to the \textit{Hilbert} space $\mathcal{H}=L_{\left[0,T\right]}^{2}$, equipped with the usual inner product:
\[
\forall x,y\in\mathcal{H}:\left\langle x,y\right\rangle =%
{\textstyle\int\nolimits_{0}^{T}}
x(t)y(t)dt
\]

In order to lighten the notation, one will henceforth omit the integration limits of the interval $\left[  0,T\right]  $ under the integral sign. One denotes by:
\[
\mu(t)=\frac{1}{n}%
{\textstyle\sum\nolimits_{i=1}^{n}}
x_{i}(t)
\]
the function, or \textquotedblleft mean curve\textquotedblright\, frequently denoted by $\bar{x}(t)$, and by:
\[
\sigma(s,t)=\frac{1}{n}%
{\textstyle\sum\nolimits_{i=1}^{n}}
(x_{i}(t)-\bar{x}(t))(x_{i}(s)-\bar{x}(s))
\]
the covariance function, also called the autocovariance function. Moreover, for all $i=1,2,...,n$, the graphical representation of $x_{i}(t)$ will be referred to as \textquotedblleft curve $x_{i}$\textquotedblright\, and one will denote by $N=\{x_{1},x_{2},...,x_{n}\}$ the set of curves, commonly called the cloud of curves.

\bigskip\ 

In the following, one will assume, as usual in practice, that the family of
curves is centered $(\bar{x}(t)=0)$ and therefore that the autocovariance function is expressed in
the simplified form:
\[
\sigma(s,t)=\frac{1}{n}%
{\textstyle\sum\nolimits_{i=1}^{n}}
x_{i}(t)x_{i}(s)
\]
which is nothing else than the empirical version of the autocovariance function
$\Gamma(s,t)=\mathbb{E}\left[  X(t)X(s)\right]  $, as defined in the context of second-order stochastic processes. Let $\xi$ be any element of $\mathcal{H}$. One sets:
\[
y_{i}=\left\langle \xi,x_{i}\right\rangle =%
{\textstyle\int}
\xi(t)x_{i}(t)dt
\]
for all $i=1,2,...,n$. By definition, the first principal component, summarizing the cloud of curves as well as possible according to the criterion below, is given by the function $\xi$ such that:
\[
\frac{1}{n}%
{\textstyle\sum\nolimits_{i=1}^{n}}
y_{i}^{2}%
\]
is maximized. To ensure that the solution is unique, one sets, as in the finite
case, the following normality constraint:
\[
||\xi||^{2}=%
{\textstyle\int}
\xi^{2}(t)dt=1
\]
The optimization problem is then expressed in the following form:
\[
\underset{\xi\in\mathcal{H}}{\sup}\frac{1}{n}%
{\textstyle\sum\nolimits_{i=1}^{n}}
\left[
{\textstyle\int}
\xi(t)x_{i}(t)dt\right]  ^{2}\text{ \ under the constraint: }%
{\textstyle\int}
\xi^{2}(t)dt=1
\]

\bigskip

Before solving this optimization problem, a few comments are in order to
connect this to the concept of the first principal component in the finite case. First,
one notes that, due to the unit norm of the function $\xi$, inner product $y_{i}=\left\langle
\xi,x_{i}\right\rangle $ between $\xi$ and  $x_{i}$ is actually the
\textquotedblleft coordinate\textquotedblright\ of the orthogonal projection of $x_{i}$ onto $\xi$.

Secondly, if in finite dimension, $N=\{x_{i},p_{i}\}$ denotes a cloud, assumed to be centered $(\Sigma_{i=1}^{n}p_{i}x_{i}=0)$, of $n$
points or observations $x_{i}$ of a vector in $\mathbb{R}^{p}$,  assigned with positive a weight $p_{i}$ such that  
$\Sigma_{i=1}^{n}p_{i}=1$, the first principal component is determined by seeking the 1-dimensional
subspace $\Delta_{u}$ of $\mathbb{R}^{p}$, with unit vector $u,$ such that the orthogonal projections $y_{i}=\left\langle u,x_{i}\right\rangle
=P_{\Delta_{u}}(x_{i}) $ of the $n$ points of $N=\{x_{i},p_{i}\}$ onto the support $\Delta_{u}$ of $u$, are as dispersed as large as possible, i.e., such that:
\[%
{\textstyle\sum\nolimits_{i=1}^{n}}
p_{i}(P_{\Delta_{u}}(x_{i}))^{2}=%
{\textstyle\sum\nolimits_{i=1}^{n}}
p_{i}y_{i}^{2}%
\]
is maximized. If, as it is often the case in practical situations, the weights are all equal to $\frac{1}{n}$, one easily recognizes the criterion used for finding the first principal curve. If one
keepss in mind that one of the objectives of classical principal component analysis is
a reduction of the dimension of the ambient space and that the summary of the
observations is obtained by orthogonally projecting the observations onto this
reduced subspace, then the first principal component is the unique 1-dimensional subspace $\Delta_{u}$ of $\mathbb{R}^{p}$, such
that the orthogonal projections of the elements of $N$ onto $\Delta_{u}$ are as dispersed as possible,
which is equivalent to saying that among all one-dimensional summaries of $N$, the one corresponding to the first principal component is, in some sense, the most \textquotedblleft readable\textquotedblright. At this point, it is difficult not to mention an obvious analogy with physics and the dynamics of a point mass system, since if $N=\{x_{i},p_{i}\}$ denotes such a system in $\mathbb{R}^{p}$, then the first principal axis of inertia of this system of points is, by definition, the
axis $\Delta_{u}$ passing through the center of gravity of $N$ for which the inertia of $N$ around it is
minimized, or equivalently the axis $\Delta_{u}$ passing through the center of gravity of $N$ for
which the inertia of $N$ around the orthogonal complement $\Delta_{u}^{\perp}$ of
$\Delta_{u}$ is maximized, due to the fact that the sum of these two inertia is constant and equal to the inertia of $N$
around its center of gravity. As it can be checked, the inertia
$I_{\Delta_{u}^{\perp}}(N)$ of $N$ around $\Delta_{u}^{\perp}$ is given by:

\[
I_{\Delta_{u}^{\perp}}(N)=%
{\textstyle\sum\nolimits_{i=1}^{n}}
p_{i}y_{i}^{2}%
\]
where for all $i=1,2,...,n$, $y_{i}$ is the orthogonal projection of
$x_{i}$ onto $\Delta_{u}$. It is clear that these comments
and interpretations apply equally to curves since the latter are identified by their
coordinates in a chosen basis.

\bigskip

Returning to the main problem, namely the search for the element $\xi\in\mathcal{H}$ solving the optimization problem:

\[
\underset{\xi\in\mathcal{H}}{\sup}\frac{1}{n}%
{\textstyle\sum\nolimits_{i=1}^{n}}
\left[
{\textstyle\int}
\xi(t)x_{i}(t)dt\right]  ^{2}\text{ \ under the constraint : }%
{\textstyle\int}
\xi^{2}(t)dt=1\text{ \ , \ }%
\]
this falls naturally within the framework of the calculus of variations, where the
functional $J(\xi)$ to be optimized and the constraint functional $K(\xi)$ are expressed respectively by:
\[
J(\xi)=\frac{1}{n}%
{\textstyle\sum\nolimits_{i=1}^{n}}
\left[
{\textstyle\int}
\xi(t)x_{i}(t)dt\right]  ^{2}%
\]
and:
\[
K(\xi)=%
{\textstyle\int}
\xi^{2}(t)dt
\]
For an introduction to the calculus of variations, as well as for the remainder of this
presentation, one can consult the manual by D.R. Smith (\cite{smi}).

\bigskip

Let $\Delta\xi$ be any element of  $\mathcal{H}$. One recalls that, in the sense of the \textit{G\^{a}teaux} derivative,
the variations $\delta J(\xi;\Delta\xi)$ and $\delta K(\xi;\Delta\xi)$ in the direction $\Delta\xi$, of the functionals $J(\xi)$ and $K(\xi)$ are given by:
\[
\delta J(\xi;\Delta\xi)=\left.  \frac{d}{d\epsilon}J(\xi+\varepsilon\Delta
\xi)\right\vert _{\varepsilon=0}%
\]
and by:
\[
\delta K(\xi;\Delta\xi)=\left.  \frac{d}{d\epsilon}K(\xi+\varepsilon\Delta
\xi)\right\vert _{\varepsilon=0}%
\]
After some elementary calculations, one easily obtains:
\[
\delta J(\xi;\Delta\xi)=\frac{2}{n}%
{\textstyle\sum\nolimits_{i=1}^{n}}
\left[
{\textstyle\int}
\xi(t)x_{i}(t)dt%
{\textstyle\int}
\Delta\xi(t)x_{i}(t)dt\right]
\]
and:
\[
\delta K(\xi;\Delta\xi)=2%
{\textstyle\int}
\xi(t)\Delta\xi(t)dt
\]
By virtue of the \textit{Euler-Lagrange} multiplier theorem, one knows that if $\xi^{\ast}$ is a candidate
for a local extremum in the domain $K=k$ (in the present case $k=1$), and if the functionals $J $
and $K$ admit weakly continuous variations in the neighborhood of
$\xi^{\ast}$, which can be easily ensured here, then at least one of the two following properties is satisfied:

\bigskip

i) The variation of $K$ at $\xi^{\ast}$ is identically zero for all
$\Delta\xi$ belonging to $\mathcal{H}$. In other words:%
\[
\delta K(\xi^{\ast};\Delta\xi)=0\text{ }\forall \Delta\xi\in\mathcal{H}%
\]

\bigskip

ii) The variation of $J$ at $\xi^{\ast}$ is proportional to the variation of $K$ at $\xi^{\ast}$, meaning there exists a
real constant $\lambda$ such that:
\[
\delta J(\xi^{\ast};\Delta\xi)=\lambda\delta K(\xi^{\ast};\Delta\xi)
\]
for all elements $\Delta\xi$ in \`{a} $\mathcal{H}$. Such a constant is called an \textit{Euler-Lagrange}
multiplier.

\bigskip

In the present case one has:
\[
\delta K(\xi^{\ast};\Delta\xi)=2%
{\textstyle\int}
\xi^{\ast}(t)\Delta\xi(t)dt
\]
and this latter expression cannot be identically zero, because if one sets $\Delta\xi=\xi^{\ast}$, one obtains :
\[
\delta K(\xi^{\ast};\xi^{\ast})=2%
{\textstyle\int}
\xi^{\ast}(t)^{2}dt
\]
which is strictly positive as long as $\xi^{\ast}\neq0$.
($\xi^{\ast}=0$ is inadmissible since this function does not
satisfy the prescribed constraint). Therefore, after simplification, one has:
\[
\frac{1}{n}%
{\textstyle\sum\nolimits_{i=1}^{n}}
\left[
{\textstyle\int}
\xi^{\ast}(t)x_{i}(t)dt%
{\textstyle\int}
\Delta\xi(t)x_{i}(t)dt\right]  =\lambda%
{\textstyle\int}
\xi^{\ast}(t)\Delta\xi(t)dt
\]
Now, since one has by definition:
\[
\sigma(s,t)=\frac{1}{n}%
{\textstyle\sum\nolimits_{i=1}^{n}}
x_{i}(t)x_{i}(s)\text{ \ , \ }%
\]
one can write the left-hand side of the previous equation in the following form:\begin{align*}
\frac{1}{n}%
{\textstyle\sum\nolimits_{i=1}^{n}}
\left[
{\textstyle\int}
\xi^{\ast}(t)x_{i}(t)dt%
{\textstyle\int}
\Delta\xi(t)x_{i}(t)dt\right]   &  =\frac{1}{n}%
{\textstyle\sum\nolimits_{i=1}^{n}}
\left[
{\textstyle\iint}
\xi^{\ast}(t)x_{i}(t)\Delta\xi(s)x_{i}(s)dtds\right] \\
&  =%
{\textstyle\iint}
\left[  \xi^{\ast}(t)\sigma(s,t)\Delta\xi(s)dtds\right] \\
&  =%
{\textstyle\int}
\left[
{\textstyle\int}
\xi^{\ast}(t)\sigma(s,t)dt\right]  \Delta\xi(s)ds
\end{align*}
It then follows that the condition:
\[
\delta J(\xi^{\ast};\Delta\xi)=\lambda\delta K(\xi^{\ast};\Delta\text{ }%
\xi)\text{ \ }\forall\Delta\xi\in\mathcal{H}%
\]
can be written as:
\[%
{\textstyle\int}
\left[
{\textstyle\int}
\xi^{\ast}(t)\sigma(s,t)dt\right]  \Delta\xi(s)ds=\lambda%
{\textstyle\int}
\xi^{\ast}(s)\Delta\xi(s)ds
\]
which, by pooling the terms, gives:
\[
\delta J(\xi^{\ast};\Delta\xi)-\lambda\delta K(\xi^{\ast};\Delta\xi)=%
{\textstyle\int}
\left[
{\textstyle\int}
\xi^{\ast}(t)\sigma(s,t)dt-\lambda\xi^{\ast}(s)\right]  \Delta\xi(s)ds
\]
Since this latter quantity must be zero for all $\Delta
\xi$ in $\mathcal{H}$, one deduces, by means of the \textit{Du Bois-Reymond } lemma (see (\cite{smi})), that:
\[%
{\textstyle\int}
\xi^{\ast}(t)\sigma(s,t)dt-\lambda\xi^{\ast}(s)=0
\]
(one can check by the way that choosing the function $\Delta\xi(s)=%
{\textstyle\int}
\xi^{\ast}(t)\sigma(s,t)dt-\lambda\xi^{\ast}(s)$, leads to a contradiction).

\bigskip 

One then considers the operator $\Gamma$ from $\mathcal{H}$ 
into itself which maps any function $x$ to the function $y=\Gamma
x$ in the following manner:
\[
y(s)=%
{\textstyle\int}
x(t)\sigma(s,t)dt
\]
where $\sigma(s,t)$ is the autocovariance function. It is known (see T. Hsing and R. Eubank (\cite{hsi}) ) that this operator, called the
\textquotedblleft covariance operator\textquotedblright, is linear, positive, self-adjoint,
and compact. It follows that the solution $\xi^{\ast}$ to the above optimization problem,
must satisfy the equation:
\[%
{\textstyle\int}
\xi^{\ast}(t)\sigma(s,t)dt=\lambda\xi^{\ast}(s)
\]
or equivalently:
\[
\Gamma\xi^{\ast}=\lambda\xi^{\ast}%
\]
which shows that $\xi^{\ast}$ is an eigenfunction of the operator $\Gamma$
associated to the eigenvalue $\lambda$. This result generalizes the one obtained in the finite-dimensional case,
where $\Gamma$ is the variance-covariance matrix $V$ given by:
\[
V=\frac{1}{n}X^{t}X
\]
where $X$ is the matrix of centered data, of size $n\times p$,
formed by the $n$ the observations of a
vector in $\mathbb{R}^{p}$. Furthermore, one has :
\begin{align*}
J(\xi^{\ast})  &  =\frac{1}{n}%
{\textstyle\sum\nolimits_{i=1}^{n}}
\left[
{\textstyle\int}
\xi^{\ast}(t)x_{i}(t)dt\right]  ^{2}\\
&  =\frac{1}{n}%
{\textstyle\sum\nolimits_{i=1}^{n}}
\left[
{\textstyle\int}
{\textstyle\int}
\xi^{\ast}(t)x_{i}(t)\xi^{\ast}(s)x_{i}(s)dtds\right] \\
&  =%
{\textstyle\int}
\xi^{\ast}(s)\left[
{\textstyle\int}
\xi^{\ast}(t)\sigma(s,t)dt\right]  ds\\
&  =\lambda%
{\textstyle\int}
\xi^{\ast2}(s)ds\\
&  =\lambda
\end{align*}
Thus, the first principal curve will be the eigenfunction $\xi_{1}$ associated with the largest eigenvalue $\lambda_{1}$ of the covariance operator $\Gamma$. Furthermore, one will have:
\[
J(\xi^{\ast})-J(\xi^{\ast}+\Delta\xi)>0
\]
for all admissible $\Delta\xi$ belonging to $\mathcal{H}$,
that is, for all $\Delta\xi$ such that $K(\xi^{\ast}+\Delta\xi)=1$.

\bigskip

Subsequent principal curves are obtained iteratively following the
\textit{Gram-Schmidt} orthonormalization process as used in the construction of an orthonormal basis of
a vector space. For example, the second principal curve $\xi_{2}$ will be the solution to the
following optimization problem:
\[
\underset{\xi\in\mathcal{H}}{\sup}\frac{1}{n}%
{\textstyle\sum\nolimits_{i=1}^{n}}
\left[
{\textstyle\int}
\xi(t)x_{i}(t)dt\right]  ^{2}\text{ \ under the constraints }||\xi
||^{2}=1\text{ \ and \ }\left\langle \xi,\xi_{1}\right\rangle =0
\]
The functional to be optimized is given as before by::
\[
J(\xi)=\frac{1}{n}%
{\textstyle\sum\nolimits_{i=1}^{n}}
\left[
{\textstyle\int}
\xi(t)x_{i}(t)dt\right]  ^{2}%
\]
while the constraint functionals $K_{1}$ and $K_{2}$ are respectively expressed in the form:
\[
K_{1}(\xi)=%
{\textstyle\int}
\xi^{2}(t)dt\text{ \ and \ }K_{2}(\xi)=%
{\textstyle\int}
\xi(t)\xi_{1}(t)dt
\]

\bigskip

To determine the solution to the optimization problem above, one recalls the \textit{Euler-Lagrange}
multiplier theorem for the case of multiple constraints, a proof of which
can be found in, among others, the manual by R. Smith(\cite{smi}).

\bigskip

One considers a finite set $K_{1},K_{2},...,K_{m}$ of functionals defined on an open set $D$ of a normed vector
space $\mathcal{X}$  and possessing variations on $D$. One denotes by:
\[
D\left[  K_{i}=k_{i}\text{ \ for \ }i=1,2,...,m\right]
\]
the non-empty subset of $D$ consisting of all vectors $x$ of $X$ that satisfy simultaneously the following constraints:
\[
K_{1}(x)=k_{1},K_{2}(x)=k_{2},...,K_{m}(x)=k_{m}%
\]
where $k_{1},k_{2},...,k_{m}$ are given real numbers. The \textit{Euler-Lagrange } multiplier theorem for multiple
constraints is stated as follows and constitutes a necessary condition for an extremum: :

\bigskip

\qquad\textbf{Theorem: } Let $J,K_{1},K_{2},...,K_{m}$ be functionals defined and possessing variations on an open set $D$
of a  normed vector space $\mathcal{X}$ and let $x^{\ast}$ be a vector corresponding to a local extremum for $J$ in $D\left[  K_{i}=k_{i}\text{
};\text{\ }i=1,2,...,m\right]  $. Suppose further that the variations of the functionals $J,K_{1},K_{2},...,K_{m}$ are weakly continuous in the neighborhood of $x^{\ast}$. Then at least, one of the two following properties must be satisfied:

\bigskip

\qquad i) the following determinant is identically zero: 
\[
\det\left\vert
\begin{tabular}
[c]{llll}%
$\delta K_{1}(x^{\ast};\Delta x_{1})$ & $\delta K_{1}(x^{\ast};\Delta x_{2})$
& $\cdots$ & $\delta K_{1}(x^{\ast};\Delta x_{m})$\\
$\delta K_{2}(x^{\ast};\Delta x_{1})$ & $\delta K_{2}(x^{\ast};\Delta x_{2})$
& $\cdots$ & $\delta K_{2}(x^{\ast};\Delta x_{m})$\\
$\vdots$ & $\vdots$ & $\vdots$ & $\vdots$\\
$\delta K_{m}(x^{\ast};\Delta x_{1})$ & $\delta K_{m}(x^{\ast};\Delta x_{2})$
& $\cdots$ & $\delta K_{m}(x^{\ast};\Delta x_{m})$%
\end{tabular}
\right\vert =0
\]
for all elements $\Delta x_{1},\Delta x_{2},...,\Delta x_{m}$
belonging to $\mathcal{X}$, or:

\bigskip

\qquad ii) the variation of $J$ at $x^{\ast}$ is a linear combination of the variations $K_{1},K_{2},...,K_{m}$ at $x^{\ast}$. More precisely, there exist constants $\mu_{1},\mu_{2},...,\mu_{m}$, known as \textit{Euler-Lagrange} multipliers, such that:
\[
\delta J(x^{\ast};\Delta x)=%
{\textstyle\sum\nolimits_{i=1}^{m}}
\mu_{i}\delta K_{i}(x^{\ast};\Delta x)
\]
for all elements $\Delta x$ in \`{a} $\mathcal{X}$.

\bigskip

Thus, regarding the second principal curve, the constraint functionals having the
expressions:
\[
K_{1}(\xi)=%
{\textstyle\int}
\xi^{2}(t)dt\text{ \ and }K_{2}(\xi)=\text{\ }%
{\textstyle\int}
\xi_{1}(t)\xi(t)dt
\]
the variations $\delta K_{1}(\xi;\Delta\xi)$ and $\delta K_{2}(\xi;\Delta\xi)$ are given respectively by:
\[
\delta K_{1}(\xi;\Delta\xi)=2%
{\textstyle\int}
\xi(t)\Delta\xi(t)dt
\]
and by :
\[
\delta K_{2}(\xi;\Delta\xi)=%
{\textstyle\int}
\xi_{1}(t)\Delta\xi(t)dt
\]
It then follows, by considering the first condition and noting that the candidate $\xi^{\ast}$ for an extremum belongs to the domain $D\left[
k_{1}=1,k_{2}=0\right]  $:
\[
\det\left\vert
\begin{tabular}
[c]{ll}%
$\delta K_{1}(\xi^{\ast};\Delta\xi_{1})$ & $\delta K_{1}(\xi^{\ast};\Delta
\xi_{2})$\\
$\delta K_{2}(\xi^{\ast};\Delta\xi_{1})$ & $\delta K_{2}(\xi^{\ast};\Delta
\xi_{2})$%
\end{tabular}
\right\vert =\det\left\vert
\begin{tabular}
[c]{ll}%
$2%
{\textstyle\int}
\xi^{\ast}(t)\Delta\xi_{1}(t)dt$ & $2%
{\textstyle\int}
\xi^{\ast}(t)\Delta\xi_{2}(t)dt$\\
$%
{\textstyle\int}
\xi_{1}(t)\Delta\xi_{1}(t)dt$ & $%
{\textstyle\int}
\xi_{1}(t)\Delta\xi_{2}(t)dt$%
\end{tabular}
\right\vert
\]
which, after calculation, yields:
\[
2%
{\textstyle\int}
\xi^{\ast}(t)\Delta\xi_{1}(t)dt%
{\textstyle\int}
\xi_{1}(t)\Delta\xi_{2}(t)dt-2%
{\textstyle\int}
\xi^{\ast}(t)\Delta\xi_{2}(t)dt%
{\textstyle\int}
\xi_{1}(t)\Delta\xi_{1}(t)dt
\]
Now, if we choose the elements $\Delta\xi_{1}$ and $\Delta\xi_{2}$ as particular values for $\xi_{1}$ and
$\xi^{\ast}$, the above determinant is equal to $-2$, which implies that the second condition must be satisfied. There exist, therefore, two real constants $\mu_{1}$ and $\mu_{2}$ such that:
\[
\delta J(x^{\ast};\Delta\xi)=\mu_{1}\delta K_{1}(\xi^{\ast};\Delta\xi)+\mu
_{1}\delta K_{1}(\xi^{\ast};\Delta\xi)
\]
which yields 
\[%
{\textstyle\sum\nolimits_{i=1}^{n}}
\left[
{\textstyle\int}
\xi^{\ast}(t)x_{i}(t)dt%
{\textstyle\int}
\Delta\xi(t)x_{i}(t)dt\right]  =\mu_{1}%
{\textstyle\int}
\xi^{\ast}(t)\Delta\xi(t)dt+\mu_{2}%
{\textstyle\int}
\xi_{1}(t)\Delta\xi(t)dt
\]
and this for every $\Delta\xi$ in \`{a} $\mathcal{H}$. 

In particular, if we choose the first principal curve $\xi_{1}$ as the function $\Delta\xi$, one can easily deduce that $\mu_{2}$
must be zero. Indeed, due to the orthonormality of the functions
$\xi^{\ast}$ and $\xi_{1}$ one has :
\[
\mu_{1}%
{\textstyle\int}
\xi^{\ast}(t)\Delta\xi(t)dt=\mu_{1}\left\langle \xi^{\ast},\xi_{1}%
\right\rangle =0\text{ \ };\text{ \ }\mu_{2}%
{\textstyle\int}
\xi_{1}^{2}(t)dt=\mu_{2}||\xi_{1}||^{2}=\mu_{2}%
\]
and:
\begin{align*}
&  \frac{1}{n}%
{\textstyle\sum\nolimits_{i=1}^{n}}
\left[
{\textstyle\int}
\xi^{\ast}(t)x_{i}(t)dt%
{\textstyle\int}
\Delta\xi(t)x_{i}(t)dt\right] \\
&  =\frac{1}{n}%
{\textstyle\sum\nolimits_{i=1}^{n}}
\left[
{\textstyle\int}
\xi^{\ast}(t)x_{i}(t)dt%
{\textstyle\int}
\xi_{1}(t)x_{i}(t)dt\right] \\
&  =%
{\textstyle\int}
\left[
{\textstyle\int}
\xi_{1}(t)\sigma(s,t)dt\right]  \xi^{\ast}(s)ds\\
&  =\lambda\left\langle \xi^{\ast},\xi_{1}\right\rangle =0
\end{align*}
Thus, the search for the second principal curve $\xi_{2}$ reduces to the previous case, and
it corresponds to the eigenfunction of the operator $\Gamma$ associated with its second
largest eigenvalue $\lambda_{2}$. Furthermore, $J(\xi_{2})=\lambda_{2}$.

\bigskip

If, as in the vast majority of applications, the operator
$\Gamma$ is of finite rank $p$, then there exist $p$ eigenvalues
$\lambda_{1}\geq\lambda_{2}\geq,...,\geq\lambda_{p}>0$ as well as $p$ principal curves $\xi_{1},\xi_{2},...,\xi_{p}$ associated with them, also called principal components, which constitute an orthonormal basis of the image $\Gamma$($\mathcal{H}$ in $\mathcal{H}$.

\subsubsection{Functions generated by a finite system of elements of $\mathcal{H}$}

One considers here the framework where the data are smoothed functions
generated by linear combinations of a finite number of linearly
independent elements of $\varphi_{1},\varphi_{2},...,\varphi_{p}$ of $\mathcal{H}$. Thus, for every
$i=1,2,...,n$, one has:
\[
x_{i}=%
{\textstyle\sum\nolimits_{j=1}^{p}}
\beta_{i,j}\varphi_{j}%
\]
By denoting $x$ and $\varphi$ as the vectors of
$\mathcal{H}^{n}$ and$\mathcal{H}^{p}$ with respective components
$x_{1},x_{2},...,x_{n}$ and $\varphi_{1},\varphi_{2},...,\varphi_{p}$ one can write the set of $n$ relationships above in matrix form:
\[
x=\beta\varphi
\]
where $\beta=\{\beta_{ij}\}$ denotes the matrix of size $n\times p$ with general term $\beta_{ij}$ for $i=1,2,...,n$ and for $j=1,2,...,p$. In the case where the data are centered, which one will assume from now on, the covariance function $\sigma(s,t)$ is given by:
\[
\sigma(s,t)=\frac{1}{n}%
{\textstyle\sum\nolimits_{i=1}^{n}}
x_{i}(s)x_{i}(t)
\]
Using momentarily the \textquotedblleft\textit{prime}%
\textquotedblright\ notation for the transposition operation instead of
the usual letter $t$ to avoid any confusion with the argument $t$, the covariance function is then expressed in the form:
\[
\sigma(s,t)=\frac{1}{n}\varphi^{\prime}(s)\beta^{\prime}\beta\varphi(t)
\]
One also considers the symmetric, positive-definite, square matrix $\Omega$, of size $p\times p$ whose
general term $\omega_{k_{1}k_{2}}$ is given by:
\[
\omega_{k_{1},k_{2}}=\left\langle \varphi_{k_{1}},\varphi_{k_{2}}\right\rangle
=%
{\textstyle\int}
\varphi_{k_{1}}(t)\varphi_{k_{2}}(t)dt
\]
or equivalently: :
\[
\Omega=%
{\textstyle\int}
\varphi(t)\varphi^{\prime}(t)dt
\]
One can easily check that in the case where the considered system
$\{\varphi_{1},\varphi_{2},...,\varphi_{p}\}$ is orthonormal, then: 
$\Omega=I$.

\bigskip

Let $\xi(t)$ be an eigenfunction associated with the eigenvalue $\lambda$
of the covariance operator $\Gamma$. In other words, the function
$\xi$ satisfies the relation:
\[%
{\textstyle\int}
\sigma(s,t)\xi(t)dt=\lambda\xi(s)
\]
One further assumes that the representation of $\xi(t)$ in the system
$\{\varphi_{1},\varphi_{2},...,\varphi_{p}\}$ is given by:
\begin{align*}
\xi(t)  &  =%
{\textstyle\sum\nolimits_{j=1}^{p}}
b_{j}\varphi_{j}(t)\\
&  =b^{\prime}\varphi(t)=\varphi^{\prime}(t)b
\end{align*}
where $b$ is the vector of components $b_{1},b_{2},...,b_{p}$.
Consequently, the expression:
\[%
{\textstyle\int}
\sigma(s,t)\xi(t)dt
\]
can be written as:
\begin{align*}%
{\textstyle\int}
\sigma(s,t)\xi(t)dt  &  =%
{\textstyle\int}
\frac{1}{n}\left[  \varphi^{\prime}(s)\beta^{\prime}\beta\varphi(t)\right]
b^{\prime}\varphi(t)dt\\
&  =\frac{1}{n}\varphi^{\prime}(s)\beta^{\prime}\beta\left[
{\textstyle\int}
\varphi(t)\varphi^{\prime}(t)dt\right]  b\\
&  =\varphi^{\prime}(s)\left[  \frac{1}{n}\beta^{\prime}\beta\Omega\right]  b
\end{align*}
It follows that the relation:
\[%
{\textstyle\int}
\sigma(s,t)\xi(t)dt=\lambda\xi(s)
\]
is expressed in matrix form as:
\[
\varphi^{\prime}(s)\left[  \frac{1}{n}\beta^{\prime}\beta\Omega\right]
b=\lambda\xi(s)
\]
and by setting $\rho=n\lambda$, (one could just as well incorporate the coefficient $n^{-1}$ into one of the
matrices $\beta^{\prime}\beta$ or
$\Omega$) one obtains:
\[
\varphi^{\prime}(s)\left[  \frac{1}{n}\beta^{\prime}\beta\Omega\right]
b=\rho\xi(s)=\rho\varphi^{\prime}(s)b
\]
It follows that one must have for all $s$:
\[
\varphi^{\prime}(s)\left[  \beta^{\prime}\beta\Omega-\rho I\right]  b=0
\]
which shows that the vector $b=(b_{1},b_{2},...,b_{p})^{\prime}$, whose components represent the coordinates of the
function $\xi$ in the system $\{\varphi_{1},\varphi_{2},...,\varphi_{p}\}$ ($b_{k}=\left\langle
\xi,\varphi_{k}\right\rangle $ for $k=1,2,...,n$), is an eigenvector of the matrix $\beta^{\prime}\beta\Omega$ associated with the eigenvalue $\rho$. Moreover, the normalization constraint $||\xi||^{2}=1$ and the orthogonality
constraint $\left\langle \xi_{1},\xi_{2}\right\rangle =0$ imply that:
\[
b^{\prime}\Omega b=1\text{ \ et \ }b_{1}^{\prime}\Omega b_{2}=0
\]
and more generally:
\[
b_{i}^{\prime}\Omega b_{j}=\delta_{i}^{j}%
\]

\bigskip

Finally, setting $u=\Omega^{\frac{1}{2}}b$, finding for the eigen elements is reduces to solving the following
problem:
\[
\Omega^{\frac{1}{2}}\beta^{\prime}\beta\Omega^{\frac{1}{2}}u=\rho u
\]
which is nothing else than finding the eigen elements of the symmetric matrix:
\[
\Omega^{\frac{1}{2}}\beta^{\prime}\beta\Omega^{\frac{1}{2}}%
\]
Since the latter is, in the vast majority of applications, positive definite, it follows
that the eigenvalues, numbering $p$, are strictly positive. The eigenfunctions or
\textquotedblleft principal components\textquotedblright, will then be obtained using the relation:
\[
b=\Omega^{-\frac{1}{2}}u
\]
It is clear that in the event that the system $\{\varphi_{1},\varphi_{2},...,\varphi_{p}\}$ is orthonormal, then
$\Omega=I$ and the problem reduces to finding the eigen elements of the matrix $\beta^{\prime}\beta$. In this regard, the choice of the system being arbitrary, just like the choice of $p $, nothing prevents us from
choosing the observed curves $x_{1},x_{2},...,x_{n}$ themselves as the reference system. In this case, one will have:
\[
\beta=I\text{ \ et \ }\Omega=\{(\omega_{i,j})_{i,j=1}^{n}\}\text{ \ o\`{u} :
}\omega_{i,j}=%
{\textstyle\int}
x_{i}(t)x_{j}(t)dt
\]
The principal curves will then be linear combinations of the observed curves and
will result from the spectral analysis of the inner product operator associated with
the matrix $\Omega$.

\subsubsection{\textbf{Principal components of a set of parametric curves in }$\mathbb{R}^{3}$}

Without loss of generality, and to simplify the presentation, one will only addresses the
case $p=3$ below, which is the most frequent and familiar case in practice and the
modifications to be made for the case $p>3$ are immediate.

\bigskip

\textbf{Examples:}

\bigskip

To fix ideas and illustrate the advantages offered by functional data analysis
compared to finite-dimensional data analysis, we consider the following examples:

\bigskip

\qquad1) In the field of health sciences, and more precisely in pediatrics, children are
monitored from birth to age $16$ years, and the values of the following variables are
recorded for each of them, as functions of time $t$:
\begin{align*}
x(t)  & :\text{ height (in cm)}\\
y(t)  & :\text{ weight (in kg)}\\
z(t)  & :\text{ head circumference (in cm)}%
\end{align*}
The measurements are unequally spaced in time, and their frequency depends on
the child's age. Thus, from $0$ to $1$ year, these measurements are generally taken
between $8$ days and $2$ weeks at birth, then every month from the $2^{nd}$
to $6^{th}$month, as well as at the $9^{th},11^{th}$ and $12^{th}$ months. From $1$ year to $6$ years the
frequency of measurements changes to twice a year (from age 2, the body mass index is also calculated), and finally, from $6$ years to $16$ years once every two years, for a total of $24$ measurements. In pediatrics, these curves in $\mathbb{R}^{3}$, are used mainly in the $0$ to $6$ age group to monitor infant development and to provide an
early diagnosis in the event of an \textquotedblleft abnormal\textquotedblright\ behavior of the curve, which reveals
itself through sudden modifications in the curvature and torsion of the curve leading to its bifurcation. This allows the pediatrician to detect warning signs of malnutrition, generally a sign of severe deficiency, or brain development problems
such as craniosynostosis (premature closure of the skull), microcephaly, or macrocephaly, all of which are linked to the behavior of the velocity vector. The following graphs illustrate these comments (head circumference is plotted on the
vertical axis):
\[%
\begin{tabular}
[c]{ll}%
\raisebox{-0cm}{\parbox[b]{6.0231cm}{\begin{center}
\fbox{\includegraphics[
trim=0.000000cm 0.000000cm -0.162652cm -1.314574cm,
height=6.0231cm,
width=6.0231cm
]%
{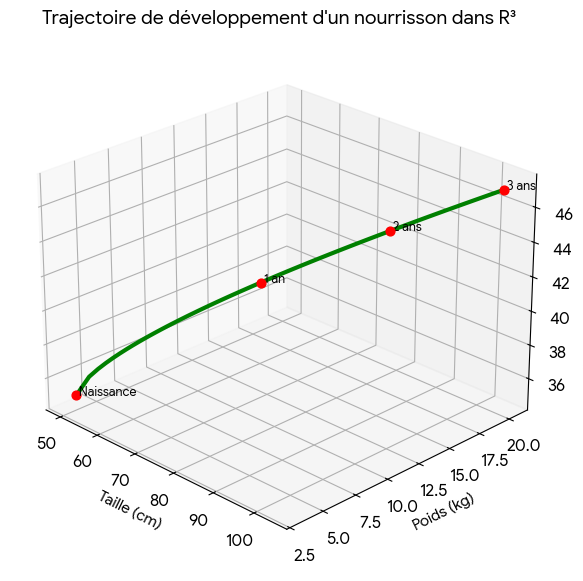}%
}\\
Figure 1: Normal development
\end{center}}}
&
\raisebox{0.0176cm}{\parbox[b]{6.0231cm}{\begin{center}
\fbox{\includegraphics[
trim=0.000000cm 0.000000cm 0.063352cm -0.090494cm,
height=6.0231cm,
width=6.0231cm
]%
{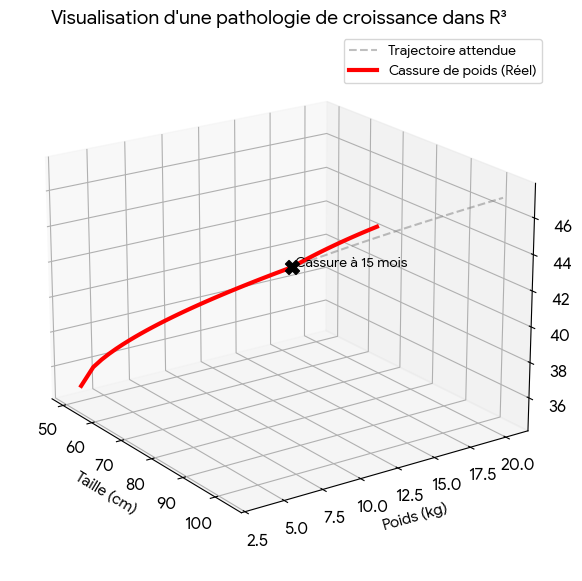}%
}\\
Figure 2: Weight bifurcation
\end{center}}}
\end{tabular}
\]%
\[%
\raisebox{-0cm}{\parbox[b]{6.0231cm}{\begin{center}
\fbox{\includegraphics[
trim=0.000000cm 0.000000cm -0.401616cm -0.002785cm,
height=6.0231cm,
width=6.0231cm
]%
{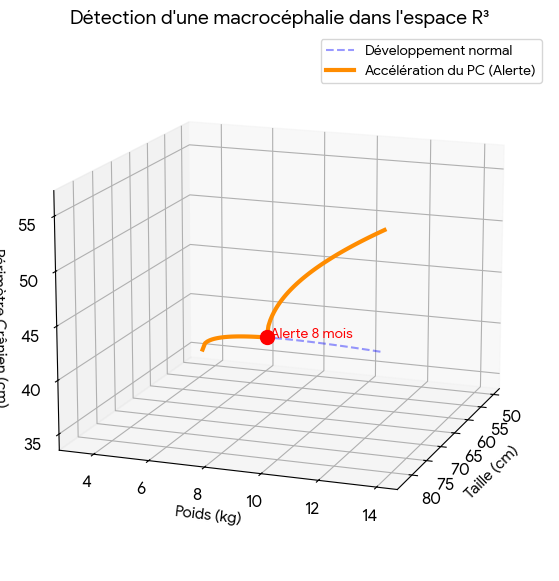}%
}\\
Figure 3: Macrocephaly
\end{center}}}
\]
It is particularly important to note here that classical multivariate data analysis does
not capture the notions of velocity, curvature, and torsion, which are essential for
the early diagnosis of growth disorders in infants and young children.

\bigskip

\qquad2) In the field of meteorology and weather forecasting, weather records are used
from a given number of stations, or observation units, scattered across a
specific territory. In an effort to combat climate change, the following variables
are taken into consideration and recorded at each station:
\begin{align*}
x(t)  &  :\text{ mean daily temperature}\\
y(t)  &  :\text{ cumulative daily precipitation amount}\\
z(t)  &  :\text{potential evaporation (amount of water evaporating from soil and plants)}\\
t  &  :\text{elapsed time in days over a year}%
\end{align*}
The variable $z(t)$ is particularly important because it accounts for sunshine hours,
mean prevailing wind speed, and humidity level at each site. Thus, each station is
described over time by a curve in $\mathbb{R}^{3}$ that meteorologists use to predict the \textquotedblleft
water stress \textquotedblright, of a region, a country, or even a continent, and whose main components
are the following:

\bigskip

\qquad\qquad i) dynamic aridity index: if the curve plunges along the $y$ axis and rises along the $x$ and $z$ axes, a severe drop in agricultural yields is anticipated.

\bigskip

\qquad\qquad ii) climate classification: curvature allows predicting whether a region is shifting
from a temperate climate to a Mediterranean or arid climate.

\bigskip

\qquad\qquad iii) flammability index: if variables $x$ and $z$ are high while variable $y$ is very low, the
risk of forest fires is maximal. The graphs below illustrate the characteristic curves
for a certain number of climate types:%
\[%
\begin{tabular}
[c]{ll}%
\raisebox{-0cm}{\parbox[b]{6.0231cm}{\begin{center}
\fbox{\includegraphics[
trim=0.000000cm 0.000000cm -0.162652cm -1.314574cm,
height=6.0231cm,
width=6.0231cm
]%
{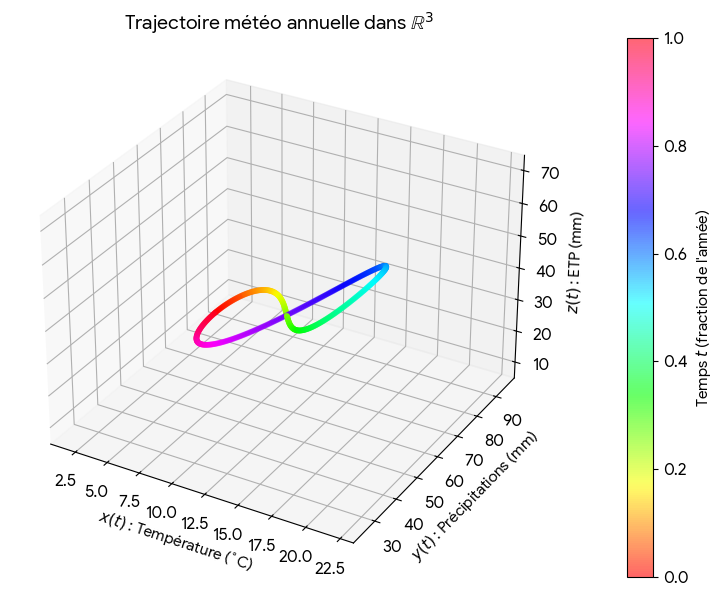}%
}\\
Figure 4: Transitional temperate climate
\end{center}}}
&
\raisebox{-0cm}{\parbox[b]{6.0231cm}{\begin{center}
\fbox{\includegraphics[
trim=0.000000cm 0.000000cm 0.053266cm -0.127454cm,
height=6.0231cm,
width=6.0231cm
]%
{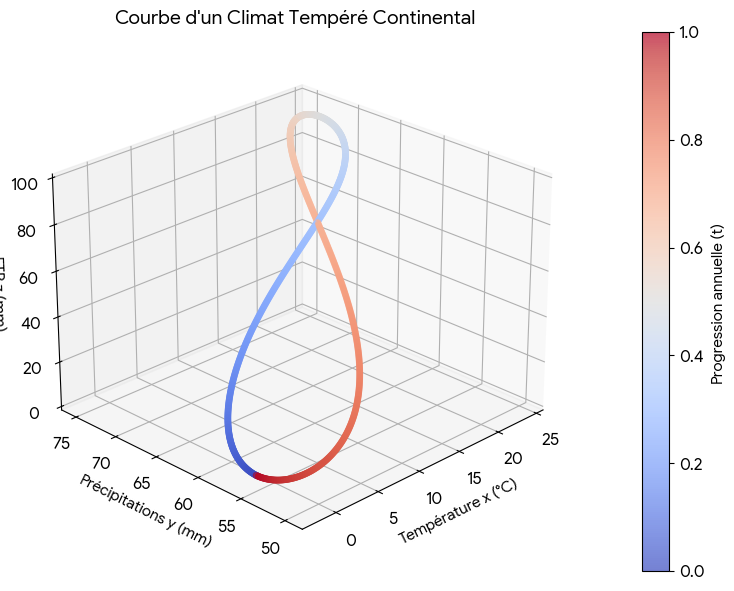}%
}\\
Figure 5: Continental temperate climate
\end{center}}}
\end{tabular}
\]%
\[%
\begin{tabular}
[c]{ll}%
\raisebox{-0cm}{\parbox[b]{6.0231cm}{\begin{center}
\fbox{\includegraphics[
trim=0.000000cm 0.000000cm -2.521970cm -0.335630cm,
height=6.0231cm,
width=6.0231cm
]%
{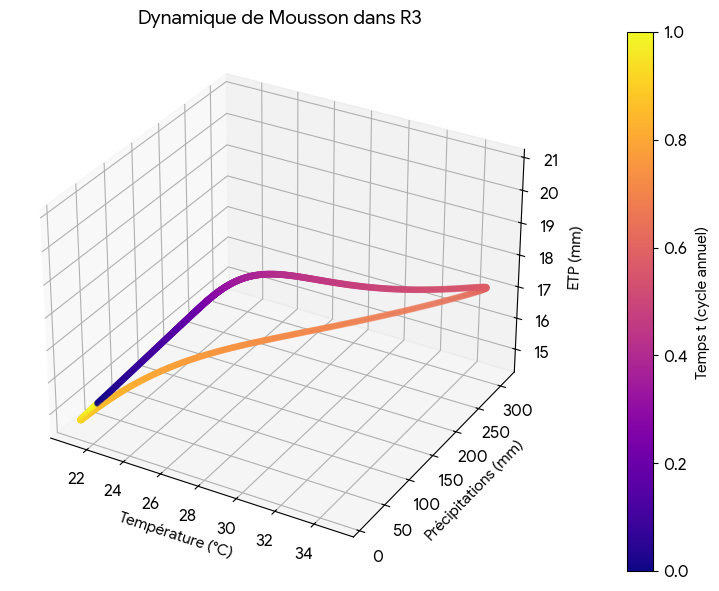}%
}\\
Figure 6: Monsoon-type climate
\end{center}}}
&
\raisebox{-0cm}{\parbox[b]{6.0231cm}{\begin{center}
\fbox{\includegraphics[
trim=0.000000cm 0.000000cm -0.066094cm -0.217782cm,
height=6.0231cm,
width=6.0231cm
]%
{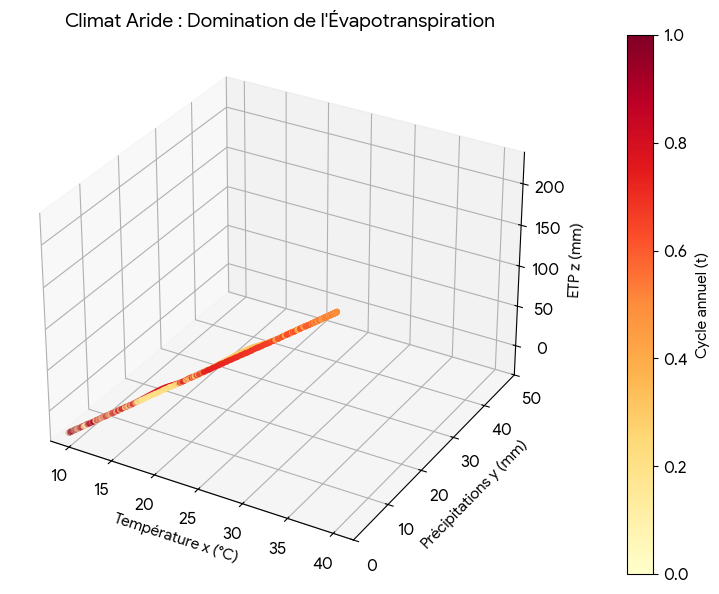}%
}\\
Figure 7: Arid climate
\end{center}}}
\end{tabular}
\]%
\[%
\begin{tabular}
[c]{ll}%
\raisebox{-0cm}{\parbox[b]{6.0231cm}{\begin{center}
\fbox{\includegraphics[
trim=0.000000cm 0.000000cm -0.298973cm -0.222320cm,
height=6.0231cm,
width=6.0231cm
]%
{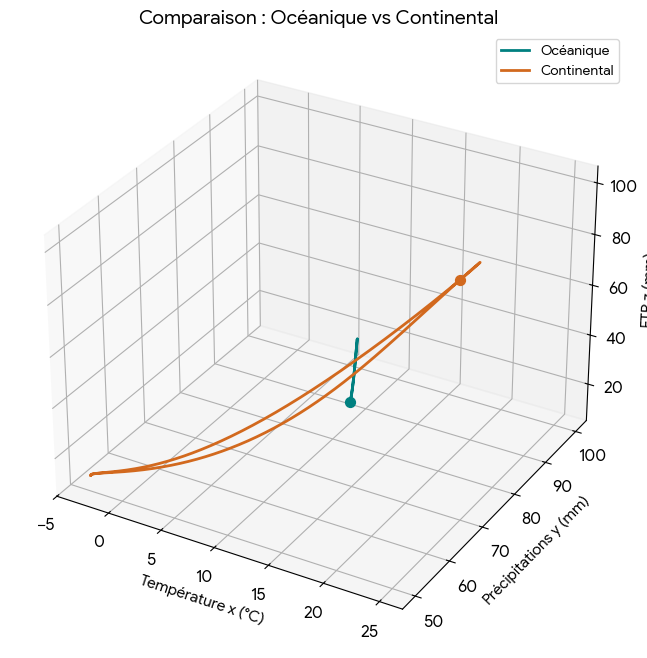}%
}\\
Figure 8: Oceanic and continental climates
\end{center}}}
&
\raisebox{0.0176cm}{\parbox[b]{6.0231cm}{\begin{center}
\fbox{\includegraphics[
trim=0.000000cm 0.000000cm -0.004273cm -0.508968cm,
height=6.0231cm,
width=6.0231cm
]%
{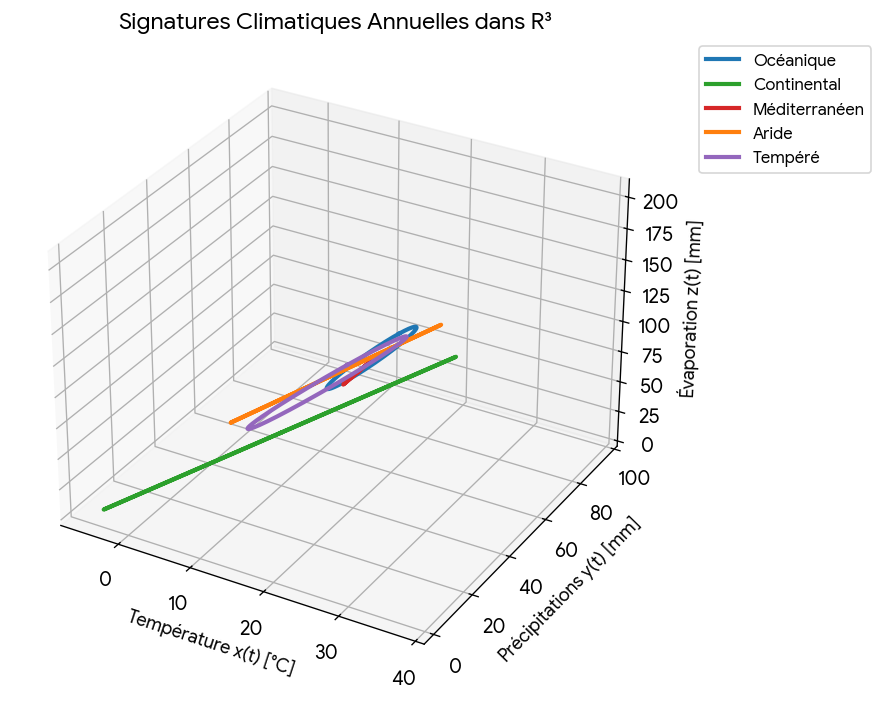}%
}\\
Figure 9: Comparison of 5 climate types
\end{center}}}
\end{tabular}
\]

\bigskip

\qquad3) To detect neurological disorders in patients, psychomotor therapists and
neurologists use a test that consists of asking the patient to draw a spiral on a
sheet of paper or a tablet. In children in particular, this test serves to evaluate
the development of gesture control (fine motor skills) and to detect neurological
disorders. Throughout the drawing process, the following variables are
measured:
\begin{align*}
t  &  :\text{elapsed time while the person draws}\\
x(t)  &  :\text{horizontal position of the stylus on the sheet (or tablet)}\\
y(t)  &  :\text{vertical position of the stylus on the sheet }\\
z(t)  &  :\text{pressure exerted by the stylus on the sheet}%
\end{align*}
Regarding children more specifically, the curve in
$\mathbb{R}^{3}$ becomes smoother and more regular as they grow. Doctors and pediatricians then analyze these curves to detect the following motor disorders in young children:

\bigskip

\qquad\qquad i) dyspraxia: if the curve in the plane $(x,y)$ is jerky and the pressure $z$ is particularly
irregular, it indicates a coordination defect.

\bigskip

\qquad\qquad ii) by analyzing the derivative (the drawing speed) as well as the curvature and
torsion of this function in $\mathbb{R}^{3}$, the specialist can detect:

\bigskip

\qquad\qquad\qquad a) a non-constant speed (jerks): the child lacks synchronism and blocks at
certain places.

\bigskip

\qquad\qquad\qquad b) visible micro-oscillations on the curve (tremor) which can take two forms:
action tremor and intention tremor (the tremor worsens as the child approaches the
end of the drawing), which is a manifestation of a cerebellar disorder.

\bigskip

\qquad\qquad\qquad c) a loss of radius control (the spiral widens too quickly), reflecting a motor
planning defect. The graphs that follow illustrate these comments in part (the pressure exerted by the stylus is plotted on the vertical axis):
\[%
\begin{tabular}
[c]{ll}%
\raisebox{-0cm}{\parbox[b]{6.0231cm}{\begin{center}
\fbox{\includegraphics[
trim=0.000000cm 0.000000cm -0.222135cm -0.246812cm,
height=6.0231cm,
width=6.0231cm
]%
{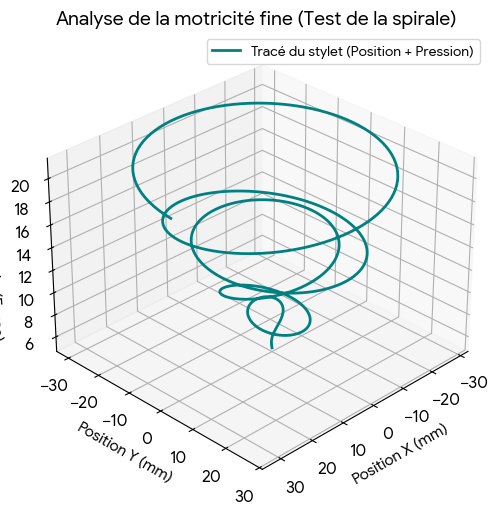}%
}\\
Figure 10: Normal curve
\end{center}}}
&
\raisebox{-0cm}{\parbox[b]{6.0231cm}{\begin{center}
\fbox{\includegraphics[
trim=0.000000cm 0.000000cm -0.391560cm -0.112770cm,
height=6.0231cm,
width=6.0231cm
]%
{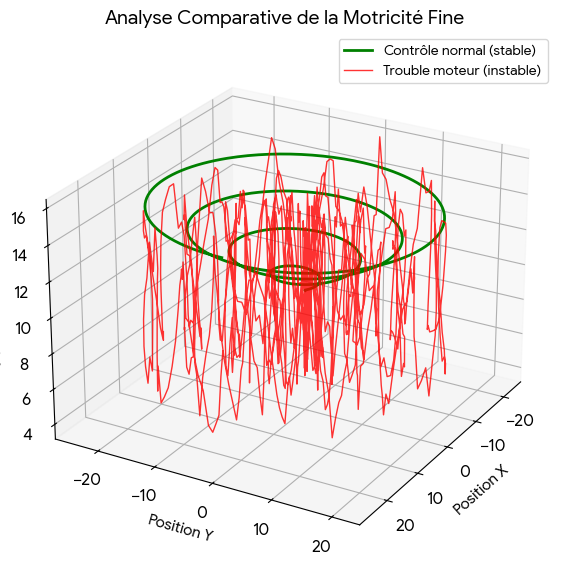}
}\\
Figure 11: Neuro-motor disorder
\end{center}}}
\end{tabular}
\]
In the right-hand figure, the physician will perceive pressure peaks on the vertical
axis, necessitating rehabilitation to teach the child to smooth this component; they
will also observe spatial irregularity between the green reference curve and the
subject's red curve, testifying to the difficulty the child's brain faces in anticipating
the spiral curve, and finally, they will note significant variations in speed, suggesting
synchronization problems with the gesture.

\bigskip

\textbf{Search for principal components of a set of parametric curves in} $\mathbb{R}^{3}$

\bigskip

In order to keep the same formalism as the one introduced in the second section, one denotes
by:
\[
\gamma=(x_{1},x_{2},x_{3})=\{\gamma(t)=(x_{1}(t),x_{2}(t),x_{3}(t)):t\in0,T\}
\]
a function of $t$ belonging, without loss of generality, to the interval $\left[  0,T\right]  $ and with vector
values in $\mathbb{R}^{3}$, whose graphical representation is called curve, usually designated
by the name: curve $\gamma$. One assumes that this curve is continuous, differentiable up to a
given order, and that it is simple and regular, which implies that the same will hold for its components $x_{1},x_{2}$ and
$x_{3}$. Furthermore, one also assumes that it belongs to the class $L_{\left[  0,T\right]
}^{2} $ on the interval $\left[  0,T\right]  $ which means that:
\[%
{\textstyle\int\nolimits_{0}^{T}}
||\gamma(t)||^{2}dt<\infty
\]
where:
\[
||\gamma(t)||^{2}=x_{1}^{2}(t)+x_{2}^{2}(t)+x_{3}^{2}(t)
\]
Recalling the theoretical framework of the second section and the notations
introduced therein, one sets:
\[
\mathcal{H}_{1}\mathcal{=H}_{2}\mathcal{=H}_{3}\mathcal{=}L_{\left[
0,T\right]  }^{2}%
\]
so that $\gamma$ is an element of the \textit{Hilbert} space $\mathcal{H}$ given by:
\[
\mathcal{H=}\left[  \times_{i=1}^{3}\mathcal{H}_{i}\right]  \mathcal{=}\left[
L_{\left[  0,T\right]  }^{2}\right]  ^{3}%
\]
One knows that in this space, the inner product between two elements $\gamma_{1}=(x_{11},x_{12},x_{13})$ and $\gamma_{2}$
$=(x_{21},x_{22},x_{23})$ is given by:
\begin{align*}
\left\langle \gamma_{1},\gamma_{2}\right\rangle _{\mathcal{H}}  &
=\left\langle x_{11},x_{21}\right\rangle _{\mathcal{H}_{1}}+\left\langle
x_{12},x_{22}\right\rangle _{\mathcal{H}_{2}}+\left\langle x_{13}%
,x_{23}\right\rangle _{\mathcal{H}_{3}}\\
&  =%
{\textstyle\int\nolimits_{0}^{T}}
\left[  x_{11}(t)x_{21}(t)+x_{12}(t)x_{22}(t)+x_{13}(t)x_{23}(t)\right]  dt
\end{align*}
or, using matrix notation:
\[
\left\langle \gamma_{1},\gamma_{2}\right\rangle _{\mathcal{H}}=%
{\textstyle\int\nolimits_{0}^{T}}
\gamma_{1}(t)^{\prime}\gamma_{2}(t)dt
\]

Having chosen, for each component, a finite number of elements $\{\xi_{ij_{i}}:i=1,2,3$ $;$ $j_{i}=1,2,...,q_{i}\}$ in each  \textit{Hilbertian} bases of $L_{\left[  0,T\right]  }^{2}$ relative to the components $x_{1},x_{2}$ et $x_{3}$, one sets:
\[
\left\{
\begin{array}
[c]{c}%
\Phi_{1}=\{\varphi_{1j_{1}}=(\xi_{1j_{1}},0,0):j_{1}=1,2,...,q_{1}\}\\
\Phi_{2}=\{\varphi_{2j_{2}}=(0,\xi_{2j_{2}},0):j_{2}=1,2,...,q_{2}\}\\
\Phi_{3}=\{\varphi_{3j_{3}}=(0,0,\xi_{3j_{3}}):j_{3}=1,2,...,q_{3}\}
\end{array}
\right.
\]

\bigskip

In this regard, the choice of bases for each component is left entirely to the
discretion of the data analyst. If, for example, it is observed that component $x_{1}$ exhibits a certain periodicity, a basis of trigonometric polynomials will be chosen for this component, whereas if component
$x_{2}$ displays behavior without any particular trend, a basis of polynomials or splines will be chosen instead, while if
component $x_{3}$ appears to behave with regular growth, the choice could be oriented toward a basis of exponentials or wavelets. Finally, in practice, and without loss of generality, one sets $q_{1}=q_{2}=q_{3}=q$.

\bigskip

For obvious practical reasons, one then considers the subspace $\widehat{\mathcal{H}}$ of $\mathcal{H}$ spanned by the
linearly independent family of vectors $\{\varphi_{ij}:i=1,2,3$ $;$ $j=1,2,...,q\}$. In other words one sets:
\[
\widehat{\mathcal{H}}=\oplus_{i=1}^{3}lin\Phi_{i}%
\]
The operator $P_{\widehat{\mathcal{H}}}$ from $\mathcal{H}$ into itself,
called \textquotedblleft orthogonal projection operator onto
$\widehat{\mathcal{H}}$\textquotedblright\, associates with any $\gamma$ of
$\mathcal{H}$ its orthogonal projection $P_{\widehat{\mathcal{H}}}(\gamma)$ on ${\widehat{\mathcal{H}}}$
whose expression is given by:
\[
P_{\widehat{\mathcal{H}}}(\gamma)=%
{\textstyle\sum\nolimits_{i=1}^{3}}
\left[
{\textstyle\sum\nolimits_{j=1}^{q}}
\left\langle \gamma,\varphi_{ij}\right\rangle \varphi_{ij}\right]
\]
where:
\[
\left\langle \gamma,\varphi_{ij}\right\rangle =\left\langle x_{i},\xi
_{ij}\right\rangle
\]
One also knows (see \cite{hsi} ) that the operator $P_{\widehat{\mathcal{H}}}$ from $\mathcal{H}$ into itself, is idempotent, linear, self-adjoint, positive definite, and of finite rank. The same properties hold for the
operator $I-P_{\widehat{\mathcal{H}}}$, from $\mathcal{H}$ into itself, which is associated with the orthogonal projection of any
element $\gamma$ of $\mathcal{H} $ onto the subspace$\widehat{\mathcal{H}}^{\perp}$. It follows that:

\[
\mathcal{H=}\widehat{\mathcal{H}}\oplus\widehat{\mathcal{H}}^{\perp}%
\]
Moreover, for any $\gamma$ belonging to $\mathcal{H}$,
$P_{\widehat{\mathcal{H}}}(\gamma)$ is the element of $\widehat{\mathcal{H}}$ closest from $\gamma$ in the sense that:
\[
P_{\widehat{\mathcal{H}}}(\gamma)=\arg\underset{\varphi\in\widehat{\mathcal{H}%
}}{\min}||\gamma-\varphi||^{2}%
\]
Naturally, the same property is true for $I-P_{\widehat{\mathcal{H}}}$ with $\widehat{\mathcal{H}}^{\perp}$.

\bigskip

In the context of data analysis, one has $n$ observations
relative to a phenomenon described by variables $x_{1},x_{2}$ and $x_{3}$ which are assumed to be continuous and
differentiable functions of a parameter $t$, such as time, and observed at values $t_{1},t_{2},...,t_{k}$ of this one. Thus, in the case of the first example, the observations correspond to the cohort of monitored children, and the variables $x_{1},x_{2},x_{3}$ and $t$ are, respectively: height, weight, head circumference, and the time at which the measurements are taken. A
classical data table could be presented in the following form:

\bigskip%

\[%
\begin{tabular}
[c]{|l|lll|l|lll|}\hline
$t\diagdown obs^{ns}$ &  & Obs. $1$ &  & $\cdots$ &  & Obs. $n$ & \\\hline
$t_{1}$ & $x_{11}(t_{1})$ & \multicolumn{1}{|l}{$x_{12}(t_{1})$} &
\multicolumn{1}{|l|}{$x_{13}(t_{1})$} & $\cdots$ & $x_{n1}(t_{1})$ &
\multicolumn{1}{|l}{$x_{n2}(t_{1})$} & \multicolumn{1}{|l|}{$x_{n3}(t_{1})$%
}\\\hline
$t_{2}$ & $x_{11}(t_{2})$ & \multicolumn{1}{|l}{$x_{12}(t_{2})$} &
\multicolumn{1}{|l|}{$x_{13}(t_{2})$} & $\cdots$ & $x_{n1}(t_{2})$ &
\multicolumn{1}{|l}{$x_{n2}(t_{2})$} & \multicolumn{1}{|l|}{$x_{n3}(t_{2})$%
}\\\hline
\multicolumn{1}{|c|}{$\vdots$} & \multicolumn{1}{|c|}{$\vdots$} &
\multicolumn{1}{|c|}{$\vdots$} & \multicolumn{1}{|c|}{$\vdots$} &
\multicolumn{1}{|c|}{$\vdots$} & \multicolumn{1}{|c|}{$\vdots$} &
\multicolumn{1}{|c|}{$\vdots$} & \multicolumn{1}{|c|}{$\vdots$}\\\hline
$t_{k}$ & $x_{11}(t_{k})$ & \multicolumn{1}{|l}{$x_{12}(t_{k})$} &
\multicolumn{1}{|l|}{$x_{13}(t_{k})$} & $\cdots$ & $x_{n1}(t_{k})$ &
\multicolumn{1}{|l}{$x_{n2}(t_{k})$} & \multicolumn{1}{|l|}{$x_{n3}(t_{k})$%
}\\\hline
\end{tabular}
\]

\bigskip

One can also represent these data in the form of $k$ clouds of $n$ points in $\mathbb{R}^{3}$ which may be visually of interest. This is the case, for example, when the cohort is composed of different groups, such as a group of girls and a group of boys.

\bigskip

Having chosen a space $\widehat{\mathcal{H}}$, the data analyst may wish, as a first step, to take into
account the continuity and differentiability properties of the phenomenon under consideration by substituting each observation
$i$ with a curve $\hat{\gamma}_{i}(t)=(\hat{x}_{i1}(t),\hat{x}_{i2}(t),\hat
{x}_{i3}(t)) $ for $i=1,2,...,n$. At this end, use will be made of the smoothing technique presented in the third section, which can be employed according to the simple least squares criterion or the weighted or penalized least squares criterion depending on the given objectives.
\bigskip

Once this stage is completed, the $n$ initial data are converted into another set of $n$ data consisting of functions, whose graphical representations in $\mathbb{R}^{3}$ correspond to the curves $\hat{\gamma}_{1}(t),\hat{\gamma}_{2}(t)...,\hat{\gamma}_{n}(t)$. On this set of curves, which is assimilated to a cloud $N$ of $n$ points
in $\mathbb{R}^{3q}$, one may wish to carry out a certain number of statistical analyses
using various models, such as linear models, regression, analysis of variance... but
also analyses usually employed in multivariate data analysis, such as principal
component analysis, canonical analysis, discriminant analysis, cluster analysis,
and many others.

\bigskip

More specifically, regarding the analysis into principal components of
the set  $\{\hat{\gamma}_{1}(t),\hat{\gamma}%
_{2}(t)...,\hat{\gamma}_{n}(t)\}$, the search for the first principal component consists in finding a function
$\gamma=(\gamma_{1},\gamma
_{2},\gamma_{3})$ of $\widehat{\mathcal{H}}$, of the form:
\[
\gamma=%
{\textstyle\sum\nolimits_{i=1}^{3}}
{\textstyle\sum\nolimits_{j=1}^{q}}
\beta_{ij}\varphi_{ij}%
\]
such that:
\[
\frac{1}{n}%
{\textstyle\sum\nolimits_{l=1}^{n}}
\left\langle \gamma,\hat{\gamma}_{l}\right\rangle ^{2}%
\]
is maximal, under the constraint:
\[
||\gamma||^{2}=1
\]

\bigskip

Now $\hat{\gamma}_{l}$, resulting from a data smoothing operation, is expressed in the chosen basis
as:
\[
\hat{\gamma}_{l}=%
{\textstyle\sum\nolimits_{i=1}^{3}}
{\textstyle\sum\nolimits_{j=1}^{q}}
\hat{\alpha}_{l,ij}\varphi_{ij}%
\]
where the coefficients $\hat{\alpha}_{l,ij}$ are known numerical values (see $3.1$). Taking into
account the expressions of $\varphi_{1j},\varphi_{2j}$ and $\varphi_{3j}$, the components of the functions $\gamma$ and
$\hat{\gamma}_{l}$ are given respectively for all $t$ in $[0,T]$ and for all $l=1,2,...,n$, by:
\[
\gamma(t)=\left(
{\textstyle\sum\nolimits_{j=1}^{q}}
\beta_{1j}\xi_{1j}(t),%
{\textstyle\sum\nolimits_{j=1}^{q}}
\beta_{2j}\xi_{2j}(t),%
{\textstyle\sum\nolimits_{j=1}^{q}}
\beta_{3j}\xi_{3j}(t)\right)
\]
and by:
\[
\hat{\gamma}_{l}(t)=\left(
{\textstyle\sum\nolimits_{j=1}^{q}}
\hat{\alpha}_{l,1j}\xi_{1j}(t),%
{\textstyle\sum\nolimits_{j=1}^{q}}
\hat{\alpha}_{l,2j}\xi_{2j}(t),%
{\textstyle\sum\nolimits_{j=1}^{q}}
\hat{\alpha}_{l,3j}\xi_{3j}(t)\right)
\]
One denotes by $\phi$ the matrix of size $3\times3q$ consisting of the vectors of the bases chosen for
each of the components. In other words, the matrix
$\phi$ has the form:
\[
\phi=\left[
\begin{tabular}
[c]{llllllllllll}%
$\xi_{11}$ & $\xi_{12}$ & $\cdots$ & $\xi_{1q}$ & $0$ & $0$ & $\cdots$ & $0$ &
$0$ & $0$ & $\cdots$ & $0$\\
$0$ & $0$ & $\cdots$ & $0$ & $\xi_{21}$ & $\xi_{22}$ & $\cdots$ & $\xi_{2q}$ &
$0$ & $0$ & $\cdots$ & $0$\\
$0$ & $0$ & $\cdots$ & $0$ & $0$ & $0$ & $\cdots$ & $0$ & $\xi_{31}$ &
$\xi_{32}$ & $\cdots$ & $\xi_{3q}$%
\end{tabular}
\right]
\]
Moreover, if for $i=1,2,3$, $\beta_{i}\ $ and $\hat{\alpha}_{l,i}$ are vectors whith $q$ components defined by:
\[
\beta_{i}=\left[
\begin{tabular}
[c]{l}%
$\beta_{i1}$\\
$\beta_{i2}$\\
$\vdots$\\
$\beta_{iq}$%
\end{tabular}
\right]  \text{ \ and by: }\hat{\alpha}_{l,i}=\left[
\begin{tabular}
[c]{l}%
$\hat{\alpha}_{l,i1}$\\
$\hat{\alpha}_{l,i2}$\\
$\vdots$\\
$\hat{\alpha}_{l,iq}$%
\end{tabular}
\right]
\]
it follows, by setting $\beta=vec(\beta_{i})$ and $\hat{\alpha}_{l}=vec(\hat
{\alpha}_{l,i})$:
\[
\gamma=\phi\beta\text{ \ et \ }\hat{\gamma}_{l}=\phi\hat{\alpha}_{l}%
\]
It should be noted that in these expressions, only the vector $\beta$ is unknown since the
numerical values of the components $\hat{\alpha}_{l}$ are determined during the smoothing
operation of the initial data, in order to replace them by a known set of
functions $\{\hat{\gamma}_{l}:$ $l=1,2,...,n\}$.

\bigskip

One will notices here the obvious analogy with classical multidimensional data
analysis in the finite case. Indeed, in this particular case, the data analyst
has $n$ observations of a random vector of $\mathbb{R}^{p}$ in the form of an matrix $X$ of size
$n\times p$ and each observation is represented by a vector of $\mathbb{R}^{p}$ equipped with its canonical orthonormal
basis.
In the present case, the data consist of the $n$ vectors $\hat{\alpha}_{l}$ of $\mathbb{R}^{3q}$, whose
components are none other than the coordinates of the functions $\hat{\gamma}_{l}$ located in the
space $\widehat{\mathcal{H}}$ equipped with the orthonormal basis
$\{\varphi_{ij}\}$ and the function $\gamma$ is represented in
$\mathbb{R}^{3q}$ by the vector $\beta$ of its components $(\beta_{ij})$
in the basis $\{\varphi_{ij}\}$.

\bigskip

This being the case, the first principal component is given by
the function $\gamma$, of unit norm, or equivalently the vector $\beta$ fulfilling the condition $||\beta||^{2}=1$,
such that the quantity:
\[
\frac{1}{n}%
{\textstyle\sum\nolimits_{l=1}^{n}}
y_{l}^{2}=\frac{1}{n}%
{\textstyle\sum\nolimits_{l=1}^{n}}
\left\langle \gamma,\hat{\gamma}_{l}\right\rangle ^{2}%
\]
is as large as possible. Noticing that $y_{l}$ corresponds to the orthogonal
projection of $\hat{\gamma}_{l}$ onto $\gamma$ or, equivalently, to the orthogonal projection of the vector $\hat{\alpha}_{l}$
onto the support $\Delta_{\beta}$ of the unit vector $\beta$, the preceding expression is equal to the inertia $I_{\Delta_{\beta}^{\perp}}(N)$ of the data cloud $N$ around the subspace $\Delta_{\beta
}^{\perp}$ orthogonal to $\Delta_{\beta}$.

\bigskip

It then follows that
: 

\begin{align*}
\left\langle \gamma,\hat{\gamma}_{l}\right\rangle  &  =%
{\textstyle\sum\nolimits_{i=1}^{3}}
\left[
{\textstyle\sum\nolimits_{j=1}^{q}}
\beta_{ij}\xi_{ij}(t),%
{\textstyle\sum\nolimits_{k=1}^{q}}
\hat{\alpha}_{l,ik}\xi_{ik}(t)\right] \\
&  =%
{\textstyle\sum\nolimits_{i=1}^{3}}
\left[
{\textstyle\sum\nolimits_{j=1}^{q}}
{\textstyle\sum\nolimits_{k=1}^{q}}
\beta_{ij}\hat{\alpha}_{l,ik}\left\langle \xi_{ij}(t),\xi_{ik}(t)\right\rangle
\right] \\
&  =%
{\textstyle\sum\nolimits_{i=1}^{3}}
\left[
{\textstyle\sum\nolimits_{j=1}^{q}}
{\textstyle\sum\nolimits_{k=1}^{q}}
\beta_{ij}\hat{\alpha}_{l,ik}%
{\textstyle\int\nolimits_{0}^{T}}
\xi_{ij}(t)\xi_{k}(t)dt\right]
\end{align*}
Now:
\[%
{\textstyle\int\nolimits_{0}^{T}}
\xi_{ij}(t)\xi_{ik}(t)dt=\delta_{j}^{k}%
\]
for $i=1,2,3$ which implies:
\begin{align*}
\left\langle \gamma,\hat{\gamma}_{l}\right\rangle  &  =%
{\textstyle\sum\nolimits_{i=1}^{3}}
{\textstyle\sum\nolimits_{j=1}^{q}}
\beta_{ij}\hat{\alpha}_{l,ij}\\
&  =\beta^{\prime}\hat{\alpha}=\left\langle \hat{\alpha},\beta\right\rangle
_{\mathbb{R}^{3q}}\text{ \ },\text{ \ }%
\end{align*}
a result that is also easily found using the matrix form, and which in the present
case gives:
\begin{align*}
\left\langle \gamma,\hat{\gamma}_{l}\right\rangle  &  =%
{\textstyle\int\nolimits_{0}^{T}}
\gamma(t)^{\prime}\hat{\gamma}_{l}(t)dt\\
&  =%
{\textstyle\int\nolimits_{0}^{T}}
\beta^{\prime}\Phi^{\prime}(t)\Phi(t)\hat{\alpha}_{l}dt\\
&  =\beta^{\prime}\left[
{\textstyle\int\nolimits_{0}^{T}}
\Phi^{\prime}(t)\Phi(t)dt\right]  \hat{\alpha}_{l}\\
&  =\beta^{\prime}\hat{\alpha}_{l}%
\end{align*}
Hence:
\[
\left\langle \gamma,\hat{\gamma}_{l}\right\rangle ^{2}=\beta^{\prime}%
\hat{\alpha}_{l}\hat{\alpha}_{l}^{\prime}\beta
\]
and therefore :
\[
\frac{1}{n}%
{\textstyle\sum\nolimits_{l=1}^{n}}
\left\langle \gamma,\hat{\gamma}_{l}\right\rangle ^{2}=\beta^{\prime}\left[
\frac{1}{n}%
{\textstyle\sum\nolimits_{l=1}^{n}}
\hat{\alpha}_{l}\hat{\alpha}_{l}^{\prime}\right]  \beta
\]

\bigskip

Returning to the usual transposition sign $t$, one notes that the matrix
\[
\frac{1}{n}%
{\textstyle\sum\nolimits_{l=1}^{n}}
\hat{\alpha}_{l}\hat{\alpha}_{l}^{t}\text{ \ },\text{ \ }%
\]
can also be written in the form:
\[
V=\frac{1}{n}X^{t}X
\]
where $X$ is a matrix of size $n\times3q$ whose row $l$ is given by:
\[
\hat{\alpha}_{l}=(\hat{\alpha}_{l,11},\hat{\alpha}_{l,12},...,\hat{\alpha
}_{l,1q},\hat{\alpha}_{l,21},\hat{\alpha}_{l,22},...,\hat{\alpha}_{l,2q}%
,\hat{\alpha}_{l,31},\hat{\alpha}_{l,32},...,\hat{\alpha}_{l,3q})
\]
and whose terms are none other than the coordinates of $\hat{\gamma}%
_{l}$ in the chosen basis of $(L_{\left[  0,T\right]  }^{2})^{3}$.

\bigskip

One can consequently conclude, following the procedure described in $4.2.2
$, that the vector $\beta_{(1)}$ associated with the first principal component, or curve, $\Gamma_{1}$, is an eigenvector of the matrix:
\[
V=\frac{1}{n}X^{t}X
\]
associated with the largest eigenvalue $\lambda_{1}$ of the latter. Furthermore, one has:
\[
\Gamma_{1}=\Phi\beta_{(1)}\text{ \ },\text{ \ }%
\]
and :
\[
\Gamma_{1}^{t}V\Gamma_{1}=\lambda_{1}%
\]
which is precisely the maximum of the objective function One further notes
that due to the relation:
\[
I_{\Delta_{\beta}^{\perp}}(N)+I_{\Delta_{\beta}}(N)=I_{0}(N)=\frac{1}{n}%
{\textstyle\sum\nolimits_{l=1}^{n}}
||\hat{\gamma}_{l}||^{2}=trV
\]
the first principal component $\beta_{(1)}$, minimizes the inertia of the cloud $N$ around the
direction $\Delta_{\beta_{(1)}}$

\bigskip

One notes that the matrix $X$ plays the role of the \textquotedblleft data matrix\textquotedblright\ as introduced in the
classical case, and the matrix $V$ is in the case where the data are centered, the empirical variance-covariance matrix $\hat{\Sigma}$. The successive principal components are then obtained according to the same procedure by introducing the usual
orthonormality constraints at each step. In practice, since the matrix $\hat{\Sigma}$ is symmetric, positive definite, it possesses $3q$ strictly positive eigenvalues, which are displayed in decreasing order: 
\[
\lambda_{1}\geq\lambda_{2}\geq...\geq\lambda_{3q-1}\geq\lambda_{3q}>0
\]
Since $tr\hat{\Sigma}$ is the sum of its eigenvalues, the following quantities:
\[
\frac{\lambda_{1}}{tr\hat{\Sigma}}\geq\frac{\lambda_{2}}{tr\hat{\Sigma}}%
\geq...\geq\frac{\lambda_{k}}{tr\hat{\Sigma}}\geq...
\]
represent the individual contributions (expressed in percentage) of the first $k$ principal components $\Gamma_{1}%
,\Gamma_{2},...,\Gamma_{k}$, and :
\[
\frac{\lambda_{1}+\lambda_{2}}{tr\hat{\Sigma}}\leq\frac{\lambda_{1}%
+\lambda_{2}+\lambda_{3}}{tr\hat{\Sigma}}\leq...\leq\frac{\lambda_{1}%
+\lambda_{2}+...+\lambda_{k}}{tr\hat{\Sigma}}%
\]
represent the cumulative contributions of the subspaces spanned by them.

\bigskip

It is known that the dual form of the optimization problem consists in expressing the
fact that the first principal curve or component is the one with respect to which the
inertia $I_{\gamma}(N)$ of the cloud $N$ of the $n$
curves $\gamma_{l}$, around $\gamma$, is as small as possible, i.e., the curve $\Gamma_{1}$ belonging to 
$\widehat{\mathcal{H}}$, such that:
\[
\Gamma_{1}=\arg\underset{\gamma\in\widehat{\mathcal{H}}:||\gamma||=1}{\min
}I_{\gamma}(N)=\arg\underset{\gamma\in\widehat{\mathcal{H}}:||\gamma
||=1}{\min}\frac{1}{n}%
{\textstyle\sum\nolimits_{l=1}^{n}}
||\left\langle \gamma,\hat{\gamma}_{l}\right\rangle \gamma-\hat{\gamma}%
_{l}||^{2}%
\]
or, equivalently, searching for the unit vector $\beta
_{(1)}$ of $\mathbb{R}^{3q}$ such that:
\[
\beta_{(1)}=\arg\underset{\beta\in\mathbb{R}^{3q}:||\beta||=1}{\min}%
I_{\Delta_{\beta}}(N)=\arg\underset{\beta\in\mathbb{R}^{3q}:||\beta||=1}{\min
}\frac{1}{n}%
{\textstyle\sum\nolimits_{l=1}^{n}}
||\left\langle \beta,\hat{\alpha}_{l}|\right\rangle \beta-\hat{\alpha}%
_{l}||^{2}%
\]
Now, the right-hand term of the preceding equality can be expressed under the form:
\begin{align*}
\frac{1}{n}%
{\textstyle\sum\nolimits_{l=1}^{n}}
||\left\langle \beta,\hat{\alpha}_{l}|\right\rangle \beta-\hat{\alpha}%
_{l}||^{2}  &  =\frac{1}{n}%
{\textstyle\sum\nolimits_{l=1}^{n}}
{\textstyle\sum\nolimits_{i=1}^{3}}
||\left\langle \beta_{i},\hat{\alpha}_{l,i}\right\rangle \beta_{i}-\hat
{\alpha}_{l,i}||^{2}\\
&  =%
{\textstyle\sum\nolimits_{i=1}^{3}}
\left[  \frac{1}{n}%
{\textstyle\sum\nolimits_{l=1}^{n}}
||\left\langle \beta_{i},\hat{\alpha}_{l,i}\right\rangle \beta_{i}-\hat
{\alpha}_{l,i}||^{2}\right]
\end{align*}
and consequently, the left-hand term of the above equality will be minimized under the constraint $||\beta||=1$, if and only if the terms on the right-hand side are themselves minimized for a set of constraints consistent with the global constraint $||\beta||=1$. Now, for each subspace $\mathcal{H}_{i}=L_{\left[  0,T\right]  }^{2},$ for $i=1,2,3$, one knows that the unit vector $\tilde{\beta}_{i}$ in $\mathbb{R}^{q}$ such that:
\[
\frac{1}{n}%
{\textstyle\sum\nolimits_{l=1}^{n}}
||\left\langle \beta_{i},\hat{\alpha}_{l,i}\right\rangle \beta_{i}-\hat
{\alpha}_{l,i}||^{2}%
\]
is minimized, is none other than the first principal component
$\tilde{\beta}_{(1),i}$ of the cloud of points in $\mathbb{R}^{q}$ associated with the curves $\hat{\gamma}_{l,i}$. Denoting then by $X_{i}$ the matrix of size $n\times q$ whose row $l$ is given by:
$\hat{\alpha}_{l,i}=(\hat{\alpha}_{l,i1},\hat{\alpha}_{l,i2},...,\hat{\alpha
}_{l,iq})$, one knows that $\tilde{\beta}_{(1),i}$ is an eigenvector of the matrix:
\[
V_{i}=\frac{1}{n}X_{i}^{t}X_{i}%
\]
associated with its largest eigenvalue $\mu_{i,1}$. Therefore one has:
\[
\beta_{i}=||\beta_{i}||\tilde{\beta}_{(1),i}
\]
The vector $\beta$ \ being known, the preceding relations illustrate the participation of the
principal components of each subspace in the principal components of the product
space.

\bigskip

\textbf{Smoothing using the penalized least squares criterion}

\bigskip

In the context of functional data analysis of a family of parametric curves in $\mathbb{R}^{3}$, it is
considered that certain characteristics of the latter are of primary importance
insofar as they provide information regarding the morphology or \textquotedblleft geometric signature \textquotedblright\ of a given curve, since they are independent of the chosen parameter. This is the case for the velocity vector and the tangent, the acceleration vector and the normal, the osculating plane and the curvature, the
torsion and the binormal, etc. All these notions, introduced moreover in $3.2$, require
the use of derivatives up to order $3$ of the considered curve. This has the
consequence that if, during the smoothing operation, one wishes to exert a certain
control over the curvature and torsion, for example, of the curves in order to avoid obtaining solutions that are too irregular and too sensitive to the initial data, it would be appropriate to use a penalized least squares criterion, involving the modules of the derivatives concerned, to a greater or lesser extent using a control parameter $\lambda$, depending on the objectives pursued and the context of the application.
Practice also shows that if one wants to exert control over a derivative of
order $r$, it is desirable to take into consideration the derivative of order $r+2$. The
literature on this subject is abundant, and one can consult with profit references (\cite{fer}), (\cite{hsi}), \cite{kok},
(\cite{ram1}), (\cite{ram2}) et (\cite{ram3}). Nevertheless, we present below
the principle of penalized least squares in an elementary case that can easily be
generalized to more complex cases.

\bigskip

Given any phenomenon whose evolution is described by a real function of a
parameter $t$, let $y_{1},y_{2},...,y_{n}$, be the measurements of the latter taken at value
$t_{1},t_{2},...,t_{n}$ of $t$. One wants, using a least squares criterion, to fit the data $y_{i}:i=1,2,...,n$, to the model:
\[
y_{i}=x(t_{i})+\epsilon_{i}%
\]
where the random variables  $\epsilon_{i}$ are centered, identically distributed, an pairwise uncorrelated
and where, by hypothesis, the function $x$ is an element of a \textit{Hilbert} space $\mathcal{H}$ of the form:

\begin{align*}
x  &  =%
{\textstyle\sum\nolimits_{k=1}^{p}}
\beta_{k}\varphi_{k}\\
&  =\beta^{t}\varphi=\varphi^{t}\beta
\end{align*}
where $\{\varphi_{1},\varphi_{2},...,\varphi_{p}\}$ is a given, finite, and orthonormal system of elements of $\mathcal{H}$ and where  $\beta_{k}=\left\langle x,\varphi_{k}\right\rangle $ for all
$k=1,2,...,p$. A function of the form:
\[
x=%
{\textstyle\sum\nolimits_{k=1}^{p}}
\beta_{k}\varphi_{k}%
\]
fitting the data $(y_{1},y_{2},...,y_{n})$ in the least squares sense is the solution $\hat{\beta}$ (if it exists), of the following optimization problem
\[
\hat{\beta}=\arg\underset{\beta\in\mathbb{R}^{p}}{\min}||y-\Phi\beta||_{D}^{2}%
\]
where $y$ represents the vector of $\mathbb{R}^{n}$ with components
$y_{1},y_{2},...,y_{n}$, where $\Phi$ is the matrix of size $n\times p$
whose the $k^{th}$ column is, for all $k=1,2,...,p$, the vector of
$\mathbb{R}^{n}$ with components $\varphi_{k}(t_{1}),\varphi_{k}(t_{2}%
),...,\varphi_{k}(t_{n})$, where $\beta=(\beta_{1},\beta_{2},...,\beta
_{p})^{t}$ is a vector of $\mathbb{R}^{p}$ and where $D$ is a square, symmetric, positive definite matrix of size$n\times n$, 
associated with the chosen inner product on $\mathbb{R}^{n}$. In the vast majority of cases, the matrix $D$ corresponds either to the identity matrix $I$ (ordinary least squares criterion)), or to a diagonal matrix formed of positive terms, called the \textquotedblleft weight matrix"\textquotedblright\ (weighted least squares criterion). Since one has:
\[
||y-\Phi\beta||_{D}^{2}=(y-\Phi\beta)^{t}D(y-\Phi\beta)
\]
the problem as stated can be written in the following matrix form:
\begin{align*}
\hat{\beta}  &  =\arg\underset{\beta\in\mathbb{R}^{p}}{\min}||y-\Phi
\beta||_{D}^{2}\\
&  =\arg\underset{\beta\in\mathbb{R}^{p}}{\min}\left[  \beta^{t}\Phi^{t}%
D\Phi\beta-2\beta^{t}\Phi^{t}Dy\right]
\end{align*}
and its solution, provided that the matrix $\Phi^{t}D\Phi$ is invertible, is expressed in the following
form, well known in linear models
\[
\hat{\beta}=\left[  \Phi^{t}D\Phi\right]  ^{-1}\Phi^{t}Dy
\]
Consequently, one will have:
\[
\hat{x}(t)=%
{\textstyle\sum\nolimits_{k=1}^{p}}
\hat{\beta}_{k}\varphi_{k}(t)
\]

One now considers the linear operator$\mathcal{D}^{m}$ from
$\mathcal{H}$ into itself defined by:
\[
\forall x\in\mathcal{H}\text{, }(\mathcal{D}^{m}x)(t)=\frac{d^{m}}{dt^{m}%
}x(t)
\]
where $m$ is any integer. This operator is called the \textquotedblleft differential operator\textquotedblright. In the frequent case in practice, where one wishes to exert control over the regularity of the function:
\[
\hat{x}(t)=%
{\textstyle\sum\nolimits_{k=1}^{p}}
\hat{\beta}_{k}\varphi_{k}(t)
\]
such as exerting control over its changes of concavity, it is customary to use a loss
function involving, in addition to the quadratic loss of least squares, the square of
the norm of the second derivative, weighted by a coefficient $\lambda\geq0$, allowing to modulate
its importance in the expression of the total loss. In other words, a low value
of:
\[
||\mathcal{D}^{2}x||^{2}=\int\left[  \frac{d^{2}}{dt^{2}}x(t)\right]  ^{2}dt
\]
testifies to the few changes of concavity of the function $x$ and if at limit, $||\mathcal{D}^{2}x||^{2}=0$, the function $x$ is a straight line. Conversely, if $||\mathcal{D}^{2}x||^{2}$ takes large values, this will testify
to the fact that the function $x$ fluctuates a lot, because it possesses a large number of changes of concavity, thus giving the impression of a rather unstable phenomenon.

\bigskip

More generally, one may be led to consider a regularity component, called a\textquotedblleft
penalty\textquotedblright\ component of the form:
\[%
{\textstyle\sum\nolimits_{m=1}^{M}}
||\mathcal{D}^{m}x||^{2}=%
{\textstyle\sum\nolimits_{m=1}^{M}}
\lambda_{m}\int\left[  \frac{d^{m}}{dt^{m}}x(t)\right]  ^{2}dst
\]
where the coefficients $\lambda_{m}$ for $m=1,2,...,M$, are positive real numbers reflecting the importance
given to each component of the penalty term.

\bigskip

As an example, one considers the following simple case where the regularity
component is the norm of the second derivative. The function $x$ \ being of the form:
\[
x=%
{\textstyle\sum\nolimits_{k=1}^{p}}
\beta_{k}\varphi_{k}%
\]
it follows that:
\[
\mathcal{D}^{2}x=%
{\textstyle\sum\nolimits_{k=1}^{p}}
\beta_{k}\mathcal{D}^{2}\varphi_{k}%
\]
and that the loss function has the expression:
\[
||y-\Phi\beta||_{D}^{2}+\lambda||%
{\textstyle\sum\nolimits_{k=1}^{p}}
\beta_{k}\mathcal{D}^{2}\varphi_{k}||^{2}%
\]
Now:
\begin{align*}
||%
{\textstyle\sum\nolimits_{k=1}^{p}}
\beta_{k}\mathcal{D}^{2}\varphi_{k}||^{2}  &  =\left\langle
{\textstyle\sum\nolimits_{k=1}^{p}}
\beta_{k}\mathcal{D}^{2}\varphi_{k},%
{\textstyle\sum\nolimits_{k=1}^{p}}
\beta_{k}\mathcal{D}^{2}\varphi_{k}\right\rangle \\
&  =%
{\textstyle\sum\nolimits_{k=1}^{p}}
{\textstyle\sum\nolimits_{l=1}^{p}}
\beta_{k}\beta_{l}\left\langle \mathcal{D}^{2}\varphi_{k},\mathcal{D}%
^{2}\varphi_{l}\right\rangle \\
&  =%
{\textstyle\sum\nolimits_{k=1}^{p}}
{\textstyle\sum\nolimits_{l=1}^{p}}
\beta_{k}\beta_{l}%
{\textstyle\int}
\mathcal{D}^{2}\varphi_{k}(s)\mathcal{D}^{2}\varphi_{l}(s)ds\\
&  =\beta^{t}\Gamma\beta
\end{align*}
where $\Gamma$ is a square matrix of size $p\times p$ whose general term $\gamma_{k,l}$ is given by:
\[
\gamma_{k,l}=\left\langle \mathcal{D}^{2}\varphi_{k},\mathcal{D}^{2}%
\varphi_{l}\right\rangle =%
{\textstyle\int}
\mathcal{D}^{2}\varphi_{k}(s)\mathcal{D}^{2}\varphi_{l}(s)ds
\]
The optimization problem is then presented in the form:
\[
\hat{\beta}=\arg\underset{\beta\in\mathbb{R}^{p}}{\min}\left[  ||y-\Phi
\beta||_{D}^{2}+\lambda||%
{\textstyle\sum\nolimits_{k=1}^{p}}
\beta_{k}\mathcal{D}^{2}\varphi_{k}||^{2}\right]
\]
and its matrix expression is given by:
\[
\hat{\beta}=\arg\underset{\beta\in\mathbb{R}^{p}}{\min}\left[  (y-\Phi
\beta)^{t}D(y-\Phi\beta)+\lambda\beta^{t}\Gamma\beta\right]
\]
Setting:
\[
\nu_{\lambda}(\beta)=(y-\Phi\beta)^{t}D(y-\Phi\beta)+\lambda\beta^{t}%
\Gamma\beta
\]
which is a real function of the $p$ variables $\beta_{1},\beta
_{2},...,\beta_{p}$, it follows :
\begin{align*}
\operatorname{grad}\nu_{\lambda}(\beta)  &  =\frac{d}{d\beta}\nu_{\lambda
}(\beta)=2\Phi^{t}D\Phi\beta-2\Phi^{t}Dy+2\lambda\Gamma\beta\\
&  =2\left[  (\Phi^{t}D\Phi+\lambda\Gamma)\beta-\Phi^{t}Dy\right]
\end{align*}
The critical value $\beta$ is then given by solving the system:
\[
\operatorname{grad}\nu_{\lambda}(\beta)=0
\]
which leads to:
\[
(\Phi^{t}D\Phi+\lambda\Gamma)\beta=\Phi^{t}Dy
\]
and, provided that the matrix $(\Phi^{t}D\Phi+\lambda\Gamma)$ is invertible, gives as a solution :
\[
\hat{\beta}=(\Phi^{t}D\Phi+\lambda\Gamma)^{-1}\Phi^{t}Dy
\]
Noting moreover that \textit{Hessian }of the function $\nu_{\lambda
}(\beta)$ has the expression:
\begin{align*}
Hess(\nu_{\lambda}(\beta))  &  =\frac{d^{2}}{d\beta^{2}}\nu_{\lambda}(\beta)\\
&  =\Phi^{t}D\Phi+\lambda\Gamma
\end{align*}
and noting that this matrix is symmetric and positive definite, one can conclude that $\hat{\beta}$ is indeed a critical point corresponding to a minimum.

\bigskip

Regarding the parameter $\lambda,$ the latter is called the
\textquotedblleft smoothing parameter\textquotedblright\ or
\textquotedblleft control parameter\textquotedblright\ and its
use allows for a trade-off between the quality of the fit, measured by the residual sum of squares,
namely:
\[
||y-\hat{y}||_{D}^{2}=||y-\Phi\hat{\beta}||_{D}^{2}\text{ \ , \ }%
\]
and the smoothness of the curve $\hat{x}(t)$, measured by the penalty component. For example, if the smoothing
parameter $\lambda$ becomes increasingly large, the loss function will increasingly favor the smoothness of the
function at the expense of the quality of the fit; in the limit, if $\lambda\rightarrow \infty$ the curve will increasingly approach the one that would be obtained using classical linear regression. Conversely, the smaller the value of $\lambda$ the
more the quality of the fit increases, but at the cost of a function containing an increasingly large number of variations and therefore changes in concavity. In practice, an intermediate value of $\lambda$ is sought that is neither too large nor too small, in order to achieve the best possible compromise between the smoothness of the curve and the quality of the fit. To this end, the techniques usually employed are resampling techniques, and more particularly simple or generalized cross-validation techniques. In this regard, one may consult, among others, the following references : P. Craven
and G. Wahba \cite{cra}, C, Gu \cite{gu}, P.J. Green and B.W. Silverman
\cite{gre}, and A. Buja, T. Hastie and R. Tibshirani \cite{buj}.

\end{document}